\documentclass[preprint2]{aastex62}

\usepackage{xspace}
\usepackage{amsmath}

\newcommand{\sn}[2]{\ensuremath{#1\times 10^{#2}}\xspace}

\newcommand{\mpsr}{\ensuremath{M_{\text{psr}}}\xspace}
\newcommand{\mco}{\ensuremath{M_{\text{co}}}\xspace}
\newcommand{\mbe}{\ensuremath{M_{\text{Be}}}\xspace}
\newcommand{\rbe}{\ensuremath{R_{\text{Be}}}\xspace}

\newcommand{\tbe}{\ensuremath{T_{\text{Be}}}\xspace}
\newcommand{\mdot}{\ensuremath{\dot{M}_{\text{w}}}\xspace}
\newcommand{\vw}{\ensuremath{v_{\text{w}}}\xspace}
\newcommand{\bestar}{MWC 148\xspace}
\newcommand{\edot}{\ensuremath{L_{\text{sd}}}\xspace}
\newcommand{\thetaic}{\ensuremath{\theta_{\text{ICS}}}\xspace}
\newcommand{\porb}{\ensuremath{P_{\text{orb}}}\xspace}
\newcommand{\rsh}{\ensuremath{R_{\text{sh}}}\xspace}
\newcommand{\rshi}[1]{\ensuremath{R_{\text{sh}, #1}}\xspace}

\newcommand{\esyn}{\ensuremath{E_{\text{syn}}}\xspace}

\newcommand{\norm}{\ensuremath{N_e}\xspace}
\newcommand{\slope}{\ensuremath{\Gamma}\xspace}
\newcommand{\normi}[1]{\ensuremath{N_{e,#1}}\xspace}
\newcommand{\slopei}[1]{\ensuremath{\Gamma_{#1}}\xspace}

\newcommand{\msun}{\ensuremath{M_{\odot}}\xspace}
\newcommand{\msunyr}{\ensuremath{M_{\odot}\text{/yr}}\xspace}
\newcommand{\rsun}{\ensuremath{R_{\odot}}\xspace}
\newcommand{\dg}{\ensuremath{^{\circ}}\xspace}

\newcommand{\src}{HESS~J0632$+$057\xspace}
\newcommand{\nustar}{\textit{NuSTAR}}

\newcommand{\fermi}{{\it Fermi-}LAT\xspace}

\def\asec{\ifmmode^{\prime\prime}\else$^{\prime\prime}$\fi}

\newcommand{\flux}{erg\,cm$^{-2}$\,s$^{-1}$\xspace}
\newcommand{\lum}{erg\,s$^{-1}$\xspace}

\graphicspath{{./}{figures/}}

\shorttitle{HESS J0632+057}
\shortauthors{NuSTAR and VERITAS}

\begin{document}

\title{Probing the Properties of the Pulsar Wind in the Gamma-Ray Binary \src with \nustar\ and VERITAS Observations}

\author{A.~Archer}\affiliation{Department of Physics and Astronomy, DePauw University, Greencastle, IN 46135-0037, USA}
\author{W.~Benbow}\affiliation{Center for Astrophysics $|$ Harvard \& Smithsonian, Cambridge, MA 02138, USA}
\author{R.~Bird}\affiliation{Department of Physics and Astronomy, University of California, Los Angeles, CA 90095, USA}
\author{A.~Brill}\affiliation{Physics Department, Columbia University, New York, NY 10027, USA}
\author{R.~Brose}\affiliation{Institute of Physics and Astronomy, University of Potsdam, 14476 Potsdam-Golm, Germany and DESY, Platanenallee 6, 15738 Zeuthen, Germany}
\author{M.~Buchovecky}\affiliation{Department of Physics and Astronomy, University of California, Los Angeles, CA 90095, USA}
\author{J.~L.~Christiansen}\affiliation{Physics Department, California Polytechnic State University, San Luis Obispo, CA 94307, USA}
\author{A.~J.~Chromey}\affiliation{Department of Physics and Astronomy, Iowa State University, Ames, IA 50011, USA}
\author{W.~Cui}\affiliation{Department of Physics and Astronomy, Purdue University, West Lafayette, IN 47907, USA and Department of Physics and Center for Astrophysics, Tsinghua University, Beijing 100084, China.}
\author{A.~Falcone}\affiliation{Department of Astronomy and Astrophysics, 525 Davey Lab, Pennsylvania State University, University Park, PA 16802, USA}
\author{Q.~Feng}\affiliation{Physics Department, Columbia University, New York, NY 10027, USA}
\author{J.~P.~Finley}\affiliation{Department of Physics and Astronomy, Purdue University, West Lafayette, IN 47907, USA}
\author{L.~Fortson}\affiliation{School of Physics and Astronomy, University of Minnesota, Minneapolis, MN 55455, USA}
\author{A.~Furniss}\affiliation{Department of Physics, California State University - East Bay, Hayward, CA 94542, USA}
\author{A.~Gent}\affiliation{School of Physics and Center for Relativistic Astrophysics, Georgia Institute of Technology, 837 State Street NW, Atlanta, GA 30332-0430}
\author{G.~H.~Gillanders}\affiliation{School of Physics, National University of Ireland Galway, University Road, Galway, Ireland}
\author{C.~Giuri}\affiliation{DESY, Platanenallee 6, 15738 Zeuthen, Germany}
\author{O.~Gueta}\affiliation{DESY, Platanenallee 6, 15738 Zeuthen, Germany}
\author{D.~Hanna}\affiliation{Physics Department, McGill University, Montreal, QC H3A 2T8, Canada}
\author{T.~Hassan}\affiliation{DESY, Platanenallee 6, 15738 Zeuthen, Germany}
\author{O.~Hervet}\affiliation{Santa Cruz Institute for Particle Physics and Department of Physics, University of California, Santa Cruz, CA 95064, USA}
\author{J.~Holder}\affiliation{Department of Physics and Astronomy and the Bartol Research Institute, University of Delaware, Newark, DE 19716, USA}
\author{G.~Hughes}\affiliation{Center for Astrophysics $|$ Harvard \& Smithsonian, Cambridge, MA 02138, USA}
\author{T.~B.~Humensky}\affiliation{Physics Department, Columbia University, New York, NY 10027, USA}
\author{P.~Kaaret}\affiliation{Department of Physics and Astronomy, University of Iowa, Van Allen Hall, Iowa City, IA 52242, USA}
\author{N.~Kelley-Hoskins}\affiliation{DESY, Platanenallee 6, 15738 Zeuthen, Germany}
\author{M.~Kertzman}\affiliation{Department of Physics and Astronomy, DePauw University, Greencastle, IN 46135-0037, USA}
\author{D.~Kieda}\affiliation{Department of Physics and Astronomy, University of Utah, Salt Lake City, UT 84112, USA}
\author{M.~Krause}\affiliation{DESY, Platanenallee 6, 15738 Zeuthen, Germany}
\author{M.~J.~Lang}\affiliation{School of Physics, National University of Ireland Galway, University Road, Galway, Ireland}
\author{G.~Maier}\affiliation{DESY, Platanenallee 6, 15738 Zeuthen, Germany}
\author{P.~Moriarty}\affiliation{School of Physics, National University of Ireland Galway, University Road, Galway, Ireland}
\author{R.~Mukherjee}\affiliation{Department of Physics and Astronomy, Barnard College, Columbia University, NY 10027, USA}
\author{D.~Nieto}\affiliation{Institute of Particle and Cosmos Physics, Universidad Complutense de Madrid, 28040 Madrid, Spain}
\author{M.~Nievas-Rosillo}\affiliation{DESY, Platanenallee 6, 15738 Zeuthen, Germany}
\author{S.~O'Brien}\affiliation{Physics Department, McGill University, Montreal, QC H3A 2T8, Canada}
\author{R.~A.~Ong}\affiliation{Department of Physics and Astronomy, University of California, Los Angeles, CA 90095, USA}
\author{A.~N.~Otte}\affiliation{School of Physics and Center for Relativistic Astrophysics, Georgia Institute of Technology, 837 State Street NW, Atlanta, GA 30332-0430}
\author{N.~Park}\affiliation{WIPAC and Department of Physics, University of Wisconsin-Madison, Madison WI, USA}
\author{A.~Petrashyk}\affiliation{Physics Department, Columbia University, New York, NY 10027, USA}
\author{K.~Pfrang}\affiliation{DESY, Platanenallee 6, 15738 Zeuthen, Germany}
\author{M.~Pohl}\affiliation{Institute of Physics and Astronomy, University of Potsdam, 14476 Potsdam-Golm, Germany and DESY, Platanenallee 6, 15738 Zeuthen, Germany}
\author{R.~R.~Prado}\affiliation{DESY, Platanenallee 6, 15738 Zeuthen, Germany}
\author{E.~Pueschel}\affiliation{DESY, Platanenallee 6, 15738 Zeuthen, Germany}
\author{J.~Quinn}\affiliation{School of Physics, University College Dublin, Belfield, Dublin 4, Ireland}
\author{K.~Ragan}\affiliation{Physics Department, McGill University, Montreal, QC H3A 2T8, Canada}
\author{P.~T.~Reynolds}\affiliation{Department of Physical Sciences, Cork Institute of Technology, Bishopstown, Cork, Ireland}
\author{D.~Ribeiro}\affiliation{Physics Department, Columbia University, New York, NY 10027, USA}
\author{G.~T.~Richards}\affiliation{Department of Physics and Astronomy and the Bartol Research Institute, University of Delaware, Newark, DE 19716, USA}
\author{E.~Roache}\affiliation{Center for Astrophysics $|$ Harvard \& Smithsonian, Cambridge, MA 02138, USA}
\author{I.~Sadeh}\affiliation{DESY, Platanenallee 6, 15738 Zeuthen, Germany}
\author{M.~Santander}\affiliation{Department of Physics and Astronomy, University of Alabama, Tuscaloosa, AL 35487, USA}
\author{S.~Schlenstedt}\affiliation{CTAO, Saupfercheckweg 1, 69117 Heidelberg, Germany}
\author{G.~H.~Sembroski}\affiliation{Department of Physics and Astronomy, Purdue University, West Lafayette, IN 47907, USA}
\author{I.~Sushch}\affiliation{Institute of Physics and Astronomy, University of Potsdam, 14476 Potsdam-Golm, Germany}
\author{A.~Weinstein}\affiliation{Department of Physics and Astronomy, Iowa State University, Ames, IA 50011, USA}
\author{P.~Wilcox}\affiliation{Department of Physics and Astronomy, University of Iowa, Van Allen Hall, Iowa City, IA 52242, USA}
\author{A.~Wilhelm}\affiliation{Institute of Physics and Astronomy, University of Potsdam, 14476 Potsdam-Golm, Germany and DESY, Platanenallee 6, 15738 Zeuthen, Germany}
\author{D.~A.~Williams}\affiliation{Santa Cruz Institute for Particle Physics and Department of Physics, University of California, Santa Cruz, CA 95064, USA}
\author{T.~J~Williamson}\affiliation{Department of Physics and Astronomy and the Bartol Research Institute, University of Delaware, Newark, DE 19716, USA}
\collaboration{(VERITAS Collaboration)}

\author{C.~J.~Hailey}\affiliation{Columbia Astrophysics Laboratory, Columbia University, New York, NY, USA}
\author{S.~Mandel}\affiliation{Columbia Astrophysics Laboratory, Columbia University, New York, NY, USA}
\author{K.~Mori}\affiliation{Columbia Astrophysics Laboratory, Columbia University, New York, NY, USA}
\collaboration{(NuSTAR Collaboration)}

\correspondingauthor{R. R. Prado}
\email{raul.prado@desy.de}

\begin{abstract}

\src is a gamma-ray binary composed of a compact object orbiting a Be star with a period of about $315$ days. Extensive X-ray and TeV gamma-ray observations have revealed a peculiar light curve containing two peaks, separated by a dip. We present the results of simultaneous observations in hard X-rays with \nustar\ and in TeV gamma-rays with VERITAS, performed in November and December 2017. These observations correspond to the orbital phases $\phi\approx0.22$ and $0.3$, where the fluxes are rising towards the first light-curve peak. A significant variation of the spectral index from 1.77$\pm$0.05 to 1.56$\pm$0.05 is observed in the X-ray data. The multi-wavelength spectral energy distributions (SED) derived from the observations are interpreted in terms of a leptonic model, in which the compact object is assumed to be a pulsar and non-thermal radiation is emitted by high-energy electrons accelerated at the shock formed by the collision between the stellar and pulsar wind. The results of the SED fitting show that our data can be consistently described within this scenario, and allow us to estimate the magnetization of the pulsar wind at the location of the shock formation. The constraints on the pulsar-wind magnetization provided by our results are shown to be consistent with those obtained from other systems.           

\end{abstract}

\keywords{gamma rays: general -- stars: individual (MWC 148) -- X-rays: binaries -- X-rays: individual (\src) }

\section{Introduction}
\label{sec:intro}


Gamma-ray binaries are systems composed of a massive star and a compact object in which the non-thermal luminosity is dominated by the gamma-ray emission ($>1$ MeV)~\citep{Dubus-2013}. Objects of this class are extremely rare, and currently, only a handful of sources have been unambiguously identified as gamma-ray binaries: PSR~B1259-63, LS~5039, LS~I$+$61~303, HESS~J0632$+$057, 1FGL~J1018.6$-$5856, LMC P3, PSR~J2032$+$4127~\citep{Paredes-2019} and 3FGL J1405.4-6119~\citep{Corbet-2019}. All of these systems host compact objects, either neutron stars or black holes, orbiting a massive star of O or B type. Their orbital periods vary substantially, spanning from 3.9~days (LS~5039) to ${\sim}50$~years (PSR~J2032$+$4127). Only PSR~B1259$-$63 and PSR~J2032$+$4127 contain known radio pulsars, while the nature of the compact object in the remaining systems is still unknown. 

Two distinct scenarios are commonly invoked to explain the origin of the non-thermal emission in gamma-ray binaries. In objects known as \textit{microquasars}, the accretion of stellar material onto the compact object produces relativistic jets where particles are accelerated to high energies and can radiate in the gamma-ray band~\citep{Mirabel-1998}. Where the system hosts a pulsar, the spin-down power of a young pulsar is transferred to high-energy particles through the acceleration of electron pairs from the pulsar wind in the termination shock formed by the interaction of the pulsar and the stellar winds~\citep{Tavani-1997}.

\src was first detected as an unidentified point-like source during H.E.S.S. observations of the Monoceros region~\citep{Aharonian-2007}. Its binary nature in association with the Be star \bestar was suggested at first and supported later by {\it XMM-Newton} observations in X-rays~\citep{Hinton-2009}. Subsequent observations by VERITAS did not lead to a detection~\citep{Acciari-2009}, indicating a substantial flux variability, which is characteristic of gamma-ray binaries. Since then, extensive observations have been performed in soft X-rays~\citep{Falcone-2010} and the TeV gamma-ray band~\citep{Aliu-2014, Aleksic-2012}. Both X-ray and gamma-ray light curves present two clear periodic peaks around $\phi\approx0.3-0.4$ and $\phi\approx0.6-0.8$ when folded to an orbital period of around $315-320$ days~\citep{Aliu-2014}. The system is uncommonly faint in the GeV band compared to other gamma-ray binaries, having been detected only recently in \fermi data~\citep{Li-2017}.  

\cite{Aragona-2010} estimated the distance of the system to be $1.1-1.7$ kpc through optical spectroscopy of the companion star \bestar. The orbital period of $\porb=321$ days was first derived by \cite{Bongiorno-2011} using X-ray light-curve data, and was later refined to be $\porb=315^{+6}_{-4}$ by~\cite{Aliu-2014}. Two distinct orbital solutions have been proposed by~\cite{Casares-2012} and, recently, by~\cite{Moritani-2018}. Both studies follow similar methodologies based on optical spectroscopic data; however, \cite{Moritani-2018} uses a more extensive dataset, which results in a set of orbital parameters entirely different from those by \cite{Casares-2012}.


The nature of the compact object in \src is still uncertain, as the implications of the two orbital solutions point in the opposite directions. \cite{Moritani-2018} suggest that the mass of the compact object must be consistent with the pulsar scenario ($\mco<2.5$ \msun) unless the inclination of the system is extremely small ($< 3$\dg). On the other hand, the solution by~\cite{Casares-2012} favors the black-hole scenario with $\mco > 2.1~ \msun$~\citep{Zamanov-2017}, being only marginally compatible with the pulsar scenario. In addition to the mass functions, indirect evidence of the pulsar scenario can only be claimed by comparing the observational features of \src to other known sources, especially PSR~B1259-63, in which the compact object is a known pulsar.

In this paper, we present simultaneous observations in X-ray by \nustar\ and TeV gamma-ray by VERITAS during November and December 2017. These observations correspond to $\phi\approx0.22$ and $0.30$, respectively, when the X-ray and TeV fluxes are rising towards the first flux peak~\citep{Aliu-2014}. The pulsar scenario is probed by fitting the spectral energy distribution (SED) to a model in which the non-thermal emission is assumed to be produced by electrons from the pulsar wind that are accelerated at the termination shock formed by the collision of the stellar and pulsar wind. The X-ray and TeV gamma-ray spectra are assumed to be produced through synchrotron radiation and inverse Compton scattering of stellar photons, respectively. Because the distance between the Be star and the compact object at the time of the observations is substantially larger than the estimated size of the circumstellar disk, the model relies on a few assumptions about the properties of the polar stellar wind, neglecting the disk component. As the result of the model fitting, we obtain a relation between the pulsar spin-down luminosity (\edot) and the pulsar-wind magnetization ($\sigma$) that is consistent with theoretical expectations. This result shows that our data can be satisfactorily described within the pulsar hypothesis.

The observations and data analysis are first described in Sec.~\ref{sec:observations} and~\ref{sec:xray}. In Sec.~\ref{sec:system} we describe the system parameters, including both orbital solutions, and define the sets of parameters adopted in this work. The model based on the pulsar-wind scenario is described in Sec.~\ref{sec:model}, and the results of the SED model fitting in Sec~\ref{sec:results}. Finally, we summarize and discuss the results in Sec~\ref{sec:conclusions}.

\section{Observations}
\label{sec:observations}

\subsection{NuSTAR}
\label{sec:nustar}

{\it NuSTAR} is composed of a pair of co-aligned high-energy X-ray focusing telescopes with focal plane modules FPMA and FPMB, which provide an imaging resolution of 18\asec\ FWHM over an energy band of 3--79 keV and a characteristic 400 eV FWHM spectral resolution at 10 keV \citep{Harrison-2013}. The absolute and relative timing accuracy of \nustar, after correcting for on-board clock drift, are 3 msec and 10 $\mu$sec, respectively \citep{Madsen-2015}. \nustar's\ broadband capabilities allow it to measure spectral properties such as photon indices with relatively high precision, with little to no dependency with interstellar medium absorption ($N_{\rm H}$).

\src\ was first observed by \nustar\ on November 22, 2017, with a 49.7~ks exposure, followed by a second observation on December 14, 2017, with a 49.6~ks exposure.  Data processing and analysis was completed using the HEASOFT (V6.22) software package, including NUSTARDAS 06Jul17$\_$v1.8.0; \nustar\ Calibration Database (CALDB) files dated August 17, 2017 were utilized.


\subsection{VERITAS}
\label{sec:veritas}

VERITAS is an array of four 12m-diameter telescopes at the Fred Lawrence Whipple Observatory~\citep{Weekes-2002}, designed to observe gamma-ray sources in the energy range between 85 GeV and 30 TeV. The observations are performed through the detection of the Cherenkov light induced by the cascade of particles produced after the interaction of the gamma-ray photon with the atmosphere. Each telescope contains a 499-pixel photomultiplier tube camera at the focal plane, adding up to a field of view of 3.5\dg~\citep{Holder-2006}. The current sensitivity of VERITAS enables detection of a source with 1\% of the Crab Nebula flux within 25 hours. The angular resolution is $<0.1\dg$ at 1 TeV~\citep{Park-2015}.

\src was observed by VERITAS for 7.4 hours between November 16 and 26, 2017, and for 6.0 hours between December 14 and 16, 2017. Observations were conducted in ``wobble'' mode in which the source was located at a 0.5\dg offset from the center of the telescope's field-of-view to allow simultaneous and symmetric evaluation of the background. The average elevation was $\approx60\dg$ for both periods, resulting in an energy threshold of $200$ GeV.

\begin{figure*}
\plottwo{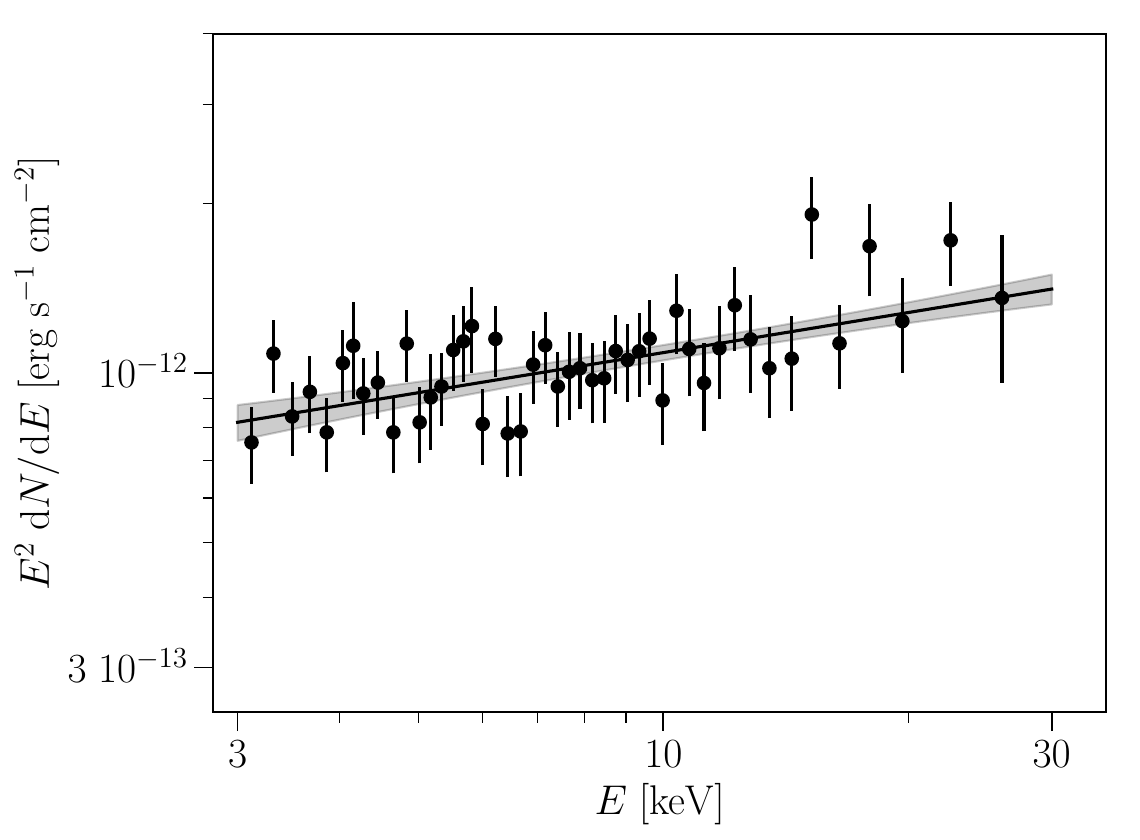}{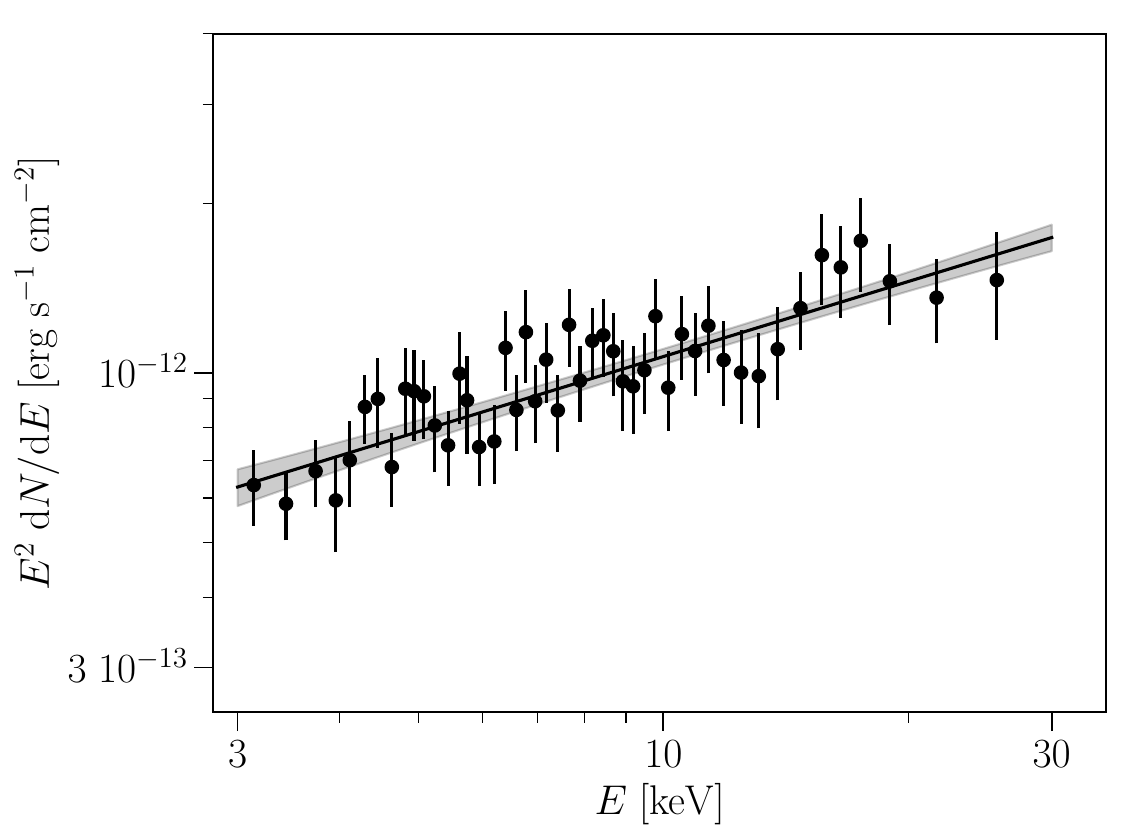}

\plottwo{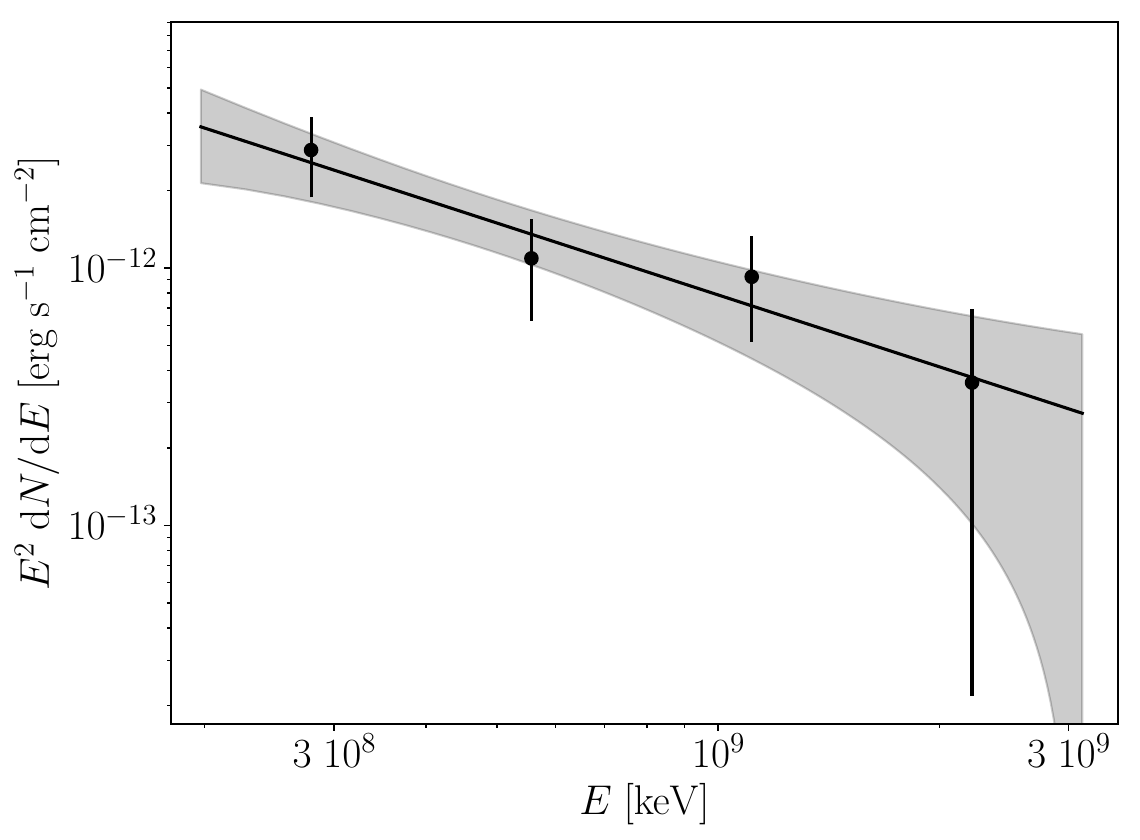}{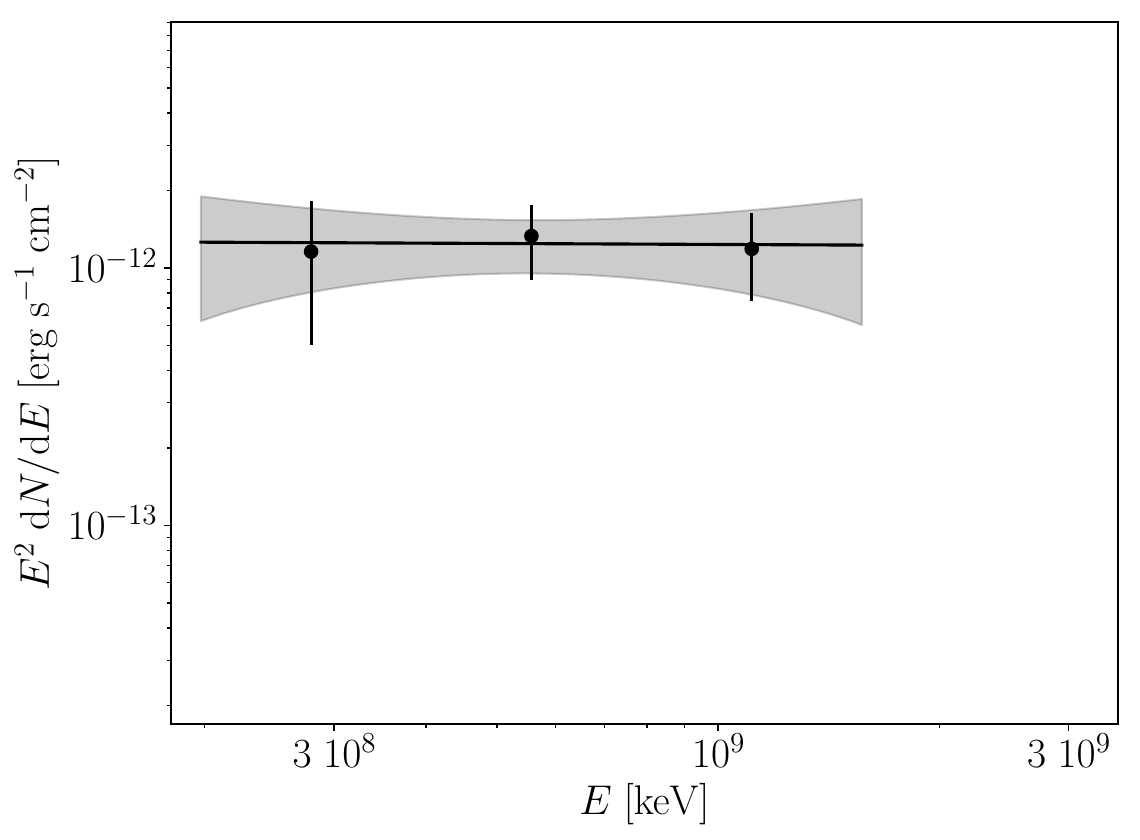}
\caption{\label{fig:obs:sed}
SED derived from \nustar\ (upper panels) and VERITAS (lower panels) observations from November (left) and December (right) 2017. The dashed lines show the result of the single power law fit and the gray band its 1$\sigma$ confidence interval.}
\end{figure*}

\section{Data Analysis}
\label{sec:xray}

\subsection{\nustar\ Spectral Analysis}

Source photons were extracted from a $r=30$\asec\ circle centered on the source position at $RA = $ 98.2472$^\circ$ and $DEC = $ 5.8015$^\circ$ (J2000), yielding a total of 2534/2468 net counts (FPMA \& FPMB combined) for the 2017 Nov/Dec observations, respectively, for a net count rate of ${\sim}0.02$ cts\,s$^{-1}$ for each detector module.  Background spectra were extracted from a rectangular source-free region on the same detector chip as the source.  We used {\tt nuproducts} to generate the response matrix and ancillary response files.

\nustar\ spectra were grouped to a minimum significance of 5$\sigma$ in each bin and fitted with the XSPEC (v12.9.1) package \citep{Arnaud-1996} using $\chi^2$ statistics. We fit \nustar\ module A and B spectra jointly in the 3--30 keV energy band, above which the background dominates. Given the previously measured  column density values ($N_{\rm H}\sim (2.1-4.7)\times10^{21}$ cm$^{-2}$) \citep{Moritani-2018}, we find that the ISM absorption is negligible above 3 keV. Therefore, \nustar\ 3--30~keV spectra allow us to determine the intrinsic continuum spectral index independently, without degeneracy with $N_{\rm H}$. 

We fit a single power-law model to the spectra.  For the November observation, the best-fit photon index was $\Gamma = 1.77\pm0.05$ ($\chi/{\text{dof}} = 0.92$ for 105 dof); the 3--30 keV flux for the unabsorbed power-law is $(2.42\pm0.13)\times10^{-12}$~\flux, and the corresponding luminosity is $(5.67\pm0.30)\times10^{32}$~\lum.  The December spectrum is significantly harder, with photon index $\Gamma = 1.56\pm0.05$ ($\chi/{\text{dof}} = 0.71$ for 102 dof), while 3--30 keV flux is $(2.45\pm0.13)\times10^{-12}$~\flux, for a luminosity of $(5.75\pm0.30)\times10^{32}$~\lum.  All uncertainties were calculated at 1$\sigma$ intervals. Luminosity values assume a distance of 1.4~kpc.

We find no significant spectral break or cutoff in the \nustar\ spectra. The \nustar\ SED derived from the spectral analysis is shown in Fig.~\ref{fig:obs:sed} (upper panels).

\subsection{\nustar\ Timing Analysis} 


After applying the barycentric correction to photon event 
files using the \nustar\ clock file, 
we extracted light curves from the same $r=30$\asec\ circular region around the source. A Bayesian block analysis yielded no time variability in the lightcurves during the \nustar\ observations. 

To produce a power density spectrum, we used \texttt{HENDRICS}, one of the modules in the \texttt{Stingray} software package \citep{Stingray-2016}. \texttt{HENDRICS} has been specifically developed for \nustar\ timing  analysis to take into account the dead time effects and observation gaps \citep{Bachetti-2015}. 
Given that \nustar\ count rates are ${\sim}0.02$~cts\,s$^{-1}$, the dead time effect is negligible. 
We binned the source lightcurves with a constant bin size $\Delta T = 2^{-6}$~sec (0.016 s) and generated power density spectra in the 3--30~keV band. 
We find that the resulting power density spectra are flat with no sign of red noise or pulsation signals during either \nustar\ observation. This is consistent with other gamma-ray binaries where the X-ray emission likely originates from an intra-binary region shocked by colliding pulsar and stellar winds~\citep{Mori-2017}. 

\subsection{VERITAS Analysis}

The analyses of VERITAS observations were performed at the position of the X-ray source XMMU~J063259.3$+$054801. The data were analyzed according to the standard procedure described by \cite{Maier-2017}. The images were first calibrated, cleaned~\citep{Acciari-2008}, and then parameterized using the Hillas criteria~\citep{Hillas-1985, Krawczynski-2006, Daniel-2008}. A stereoscopic technique that combines the orientation of the images from different telescopes was used to determine the arrival direction and core location of the gamma-ray shower. The energy reconstruction was performed by a lookup table method utilizing the impact distance, image size and zenith angle.  

A set of pre-defined cuts optimized to provide highest sensitivity for a point-like source were applied to reject background events, which are mostly composed of air showers initiated by cosmic rays. The remaining background was then estimated through the reflected region method~\citep{Fomin-1994}. The source was detected with a significance of 5.2 and 4.5$\sigma$ for the November and December 2017 observations, respectively. 

The SED derived from the VERITAS observations covers the range of $0.2-3$ TeV. In Fig.~\ref{fig:obs:sed} (lower panels), we show the SED from both observations, along with the results of a fit to a single power law. The flux in the range $0.2-3$ TeV, the corresponding luminosity, and the power law index were found to be $(3.5\pm0.8)\times 10^{-12}$~\flux, $(8.1\pm1.9)\times10^{32}$~\lum and $2.9\pm0.5$ for the November observation, and $(3.4\pm0.8)\times10^{-12}$~\flux, $(7.9\pm1.9)\times10^{32}$~\lum and $2.0\pm0.4$ for the December observation. 

\section{System Parameters and Orbital Solutions}
\label{sec:system}

Accurate knowledge of the system's geometry and the properties of the companion star are critical ingredients for any attempt at modelling gamma-ray binaries. In this section we summarize the available orbital solutions and the parameters related to the companion Be star, including the properties of the stellar wind. We also describe the sets of parameters that will be assumed for the model fitting presented in the next sections.

The properties of the companion Be star \bestar (=HD 259440) were first derived through optical spectroscopy by~\cite{Aragona-2010}. The effective temperature $T$ was found to be $27.5-30.0$ kK, the mass $13.2-19.0$~\msun and the radius $6.0-9.6$~\rsun. The distance of the system was also derived to be $1.1-1.7$ kpc. Based on these ranges of values, we assume $\tbe = 30$ kK, $\rbe=7.8$~\rsun and $d=1.4$ kpc. The mass of the star \mbe, however, will be allowed to vary within the derived range according to the mass function $f(M)$ obtained in the orbital solutions and the system's inclination $i$.   

\begin{figure*}
\plottwo{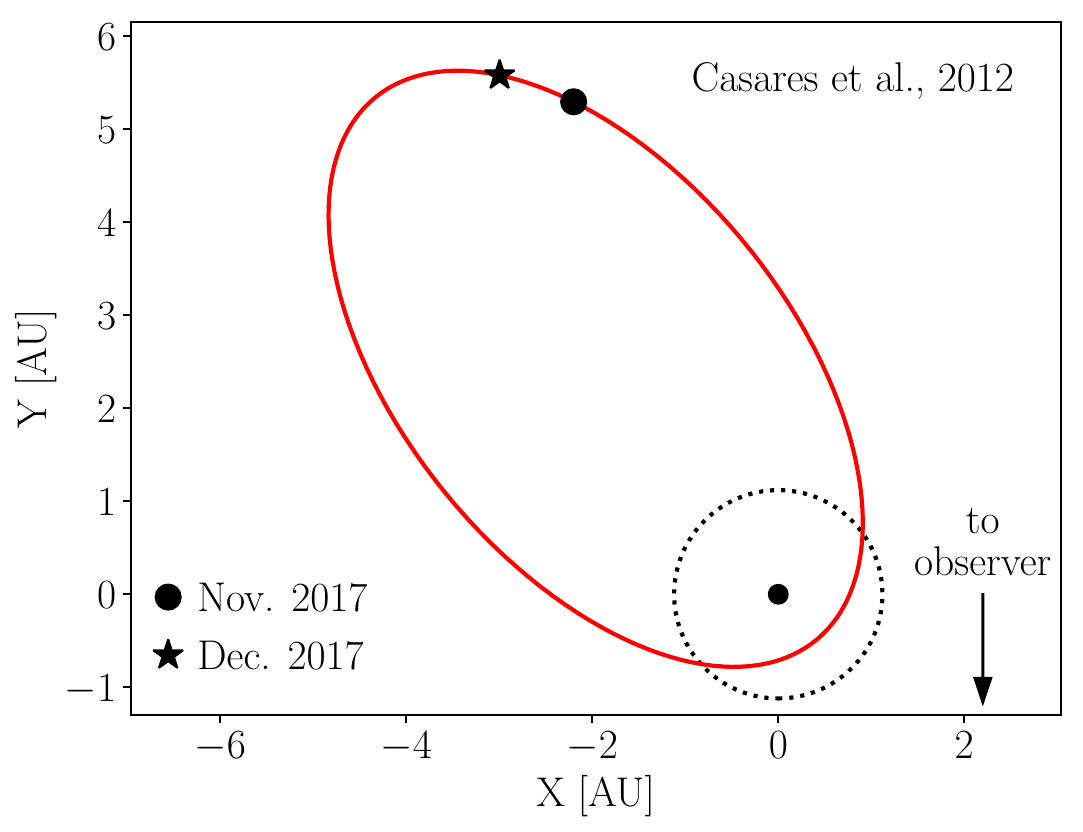}{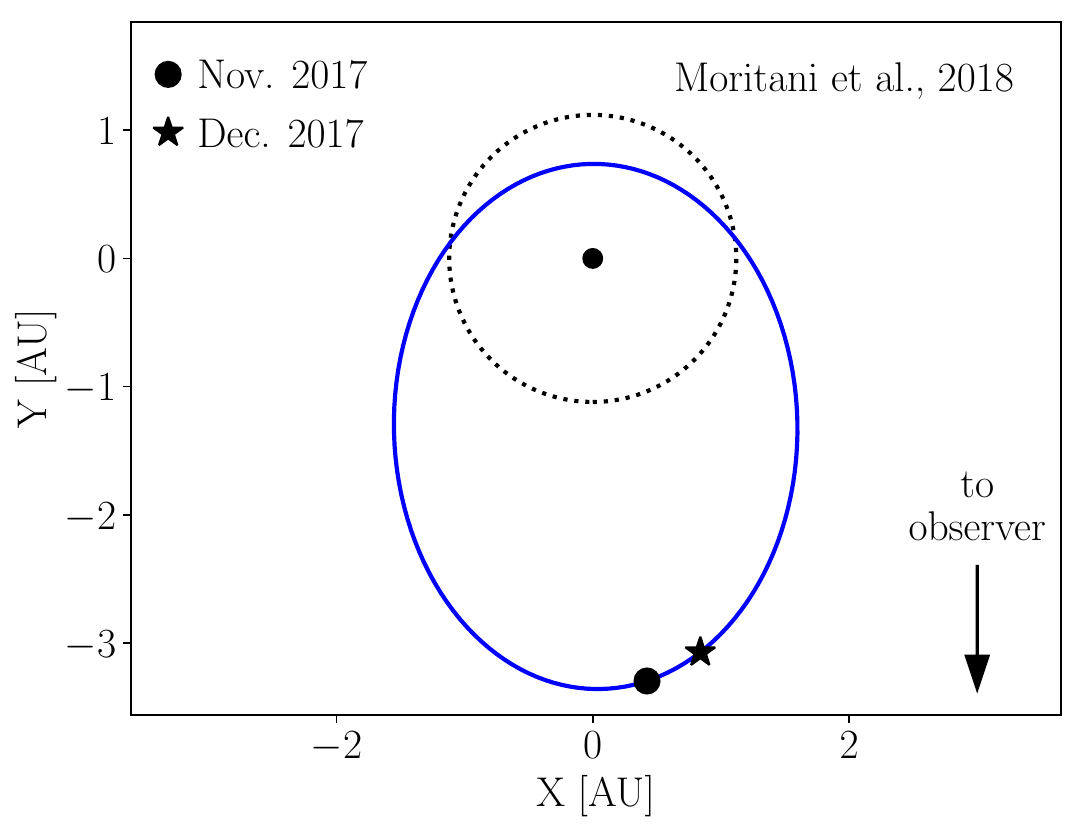}
\caption{\label{fig:system:orbits}
Illustration of the orbit of the compact object projected onto the orbital plane for~\cite{Casares-2012} (left) and~\cite{Moritani-2018} (right) solutions. The semi-major axis of the compact object ($a_2$) was calculated for the corresponding inclination ($i$) and mass function ($f(M)$) as given in Tab.~\ref{tab:system}. The locations of the compact object during the two sets of observations are indicated as black markers. The companion star is assumed to be in a fixed position and the estimated size of the circumstellar disk~\citep{Moritani-2015, Zamanov-2016} is indicated by a dashed black line.}
\end{figure*}

The first orbital solution was obtained by~\cite{Casares-2012} through spectroscopic studies of H$\alpha$ emission lines. Recently, however, \cite{Moritani-2018} proposed a distinct solution, obtained with the same methodology but with a larger dataset. For the present study, both solutions will be considered and two sets of parameters will be defined. The orbital period (\porb) of the system was assumed to be 321 days by~\cite{Casares-2012} and derived to be 313 days by~\cite{Moritani-2018}\footnote{From the two solutions obtained by~\cite{Moritani-2018}, we adopted that in which the orbital period was derived from the analysis of the X-ray light curve.}. Although \porb is correlated to the more recent orbital parameters~\citep{Ho-2017}, we assume here that the impact of varying it by a few days is negligible and we will set $\porb=315$ days for both sets of parameters, following the results obtained by~\cite{Aliu-2014}. To estimate the impact of the \porb uncertainties on the results of our studies, we will consider an arbitrary $\pm2$ days uncertainty range. The sets of parameters defined for our study are shown in Tab.~\ref{tab:system} and the orbit of the compact object is illustrated in Fig.~\ref{fig:system:orbits}. The uncertainties shown in Tab.~\ref{tab:system} were accounted for in order to probe the robustness of the model fitting. The uncertainties on \rbe and \tbe are not relevant and will be neglected.     

\begin{table*}
\centering
\caption{\label{tab:system}
System parameters of \src as used in this work. See text for the description and references. The MJD of $\phi=0$ is 54857.5 as defined by~\cite{Falcone-2010}}
\begin{tabular}{|c|c|c|}
    \hline
    & \cite{Casares-2012} & \cite{Moritani-2018} \\ \hline
    $e$ & $0.83\pm0.08$ & $0.64\pm0.29$\\
    $\omega$ (\dg) & $129\pm17$ & $271\pm29$\\
    $f$ (\msun) & 0.01 & 0.0024\\
    $i$ (\dg) & $69.5\pm10.5$ & $37\pm5$\\
    $a_2$ (AU) & $3.90^{+0.13}_{-0.22}$ & $2.13^{+0.14}_{-0.17} $\\ \hline
    \porb (d) & \multicolumn{2}{c|}{$315\pm 2$}\\
    \mpsr (\msun) & \multicolumn{2}{c|}{1.4} \\
    \mbe (\msun) & \multicolumn{2}{c|}{$13.2-19.0$} \\
    \rbe (\rsun) & \multicolumn{2}{c|}{7.8} \\
    \tbe (K) & \multicolumn{2}{c|}{30000} \\
    \vw (km/s) & \multicolumn{2}{c|}{1500} \\
    \mdot (\msunyr) & \multicolumn{2}{c|}{$10^{8.5\pm0.5}$} \\ \hline
\end{tabular}
\end{table*}

\begin{figure}
\plotone{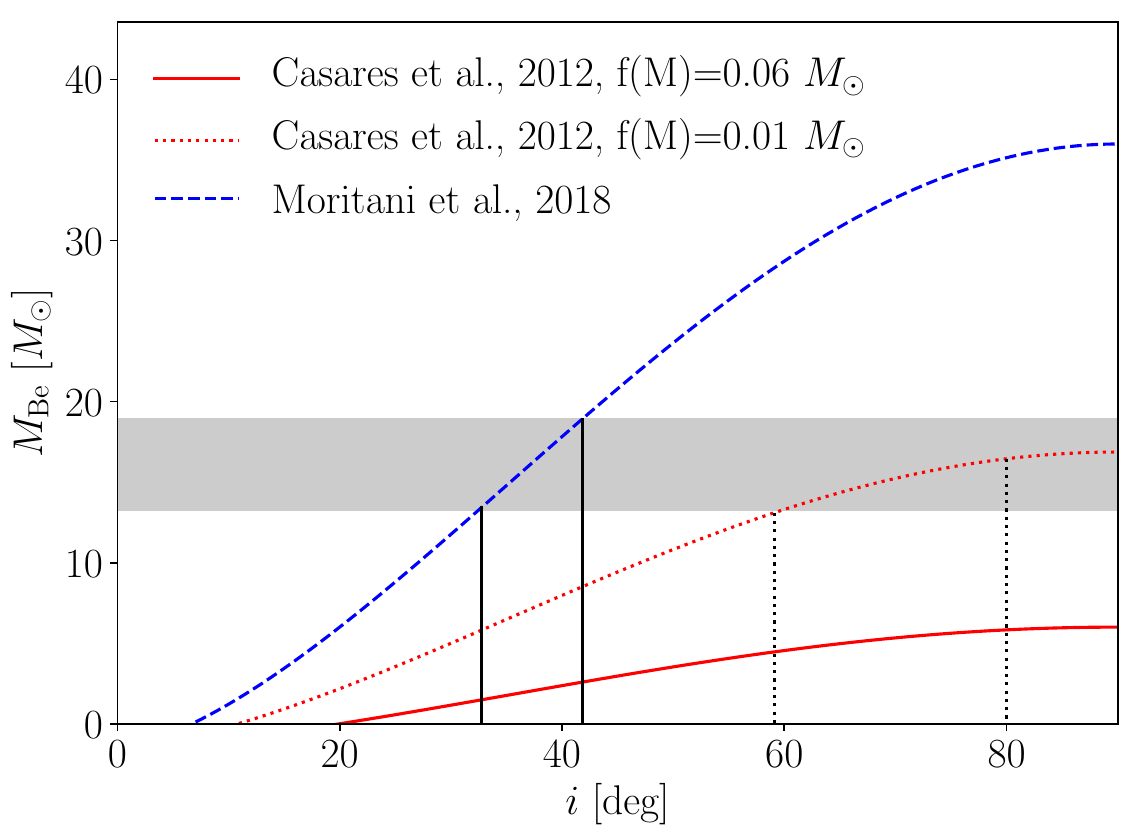}
\caption{\label{fig:system:massfunction}
The mass of the companion star (\mbe) as a function of the inclination ($i$) following the mass function given by the orbital solutions. The gray band indicates the allowed range of \mbe derived by~\cite{Aragona-2010}. The vertical solid and dotted lines indicate the limits of $i$ used in this work for the~\cite{Casares-2012} and~\cite{Moritani-2018} solutions, respectively.}
\end{figure}

The compact object is assumed to be a pulsar with $\mpsr=1.4$ \msun in the model considered here. With this assumption, the mass function $f(M)$ from a given orbital solution constrains the relation between \mbe and the inclination $i$. Thus, the allowed range of $i$ can be defined by imposing that \mbe is consistent with the range derived by~\cite{Aragona-2010}. $\mbe\;vs\;i$ is shown in Fig.~\ref{fig:system:massfunction}, from which we find that a) $32\dg<i<42\dg$ for the~\cite{Moritani-2018} solution and b) there is no value of $i$ allowed for the~\cite{Casares-2012} solution assuming the nominal value of $f(M)=0.06$, as can be seen by the solid red line. To define a set of orbital parameters based on the~\cite{Casares-2012} solution, we redefine $f(M)$ to its lower bound within the quoted uncertainties, given by $f(M)=0.06-0.05=0.01$. By doing this, we find that $i>59\dg$ becomes allowed (see the dashed red line in Fig.~\ref{fig:system:massfunction}). The upper bound of $i$ will be set to $80\dg$ following the argument given by~\cite{Casares-2012} based on the lack of observed shell lines, which would have indicated obscuration of the star by the circumstellar disk.

In the orbital parameters defined for our study, the inclination will be taken as the center of the allowed range, with the full range being used to probe the robustness of the results. Since \mbe is linked to the inclination through the mass function, $i$ and \mbe are strongly correlated. A similar relation occurs with the semi-major axis of the pulsar orbit ($a_2$), since the orbital solutions directly provide $a_1 \sin i$, where $a_1/a_2 = \mpsr/\mbe$. The uncertainties of $a_2$ indicated in Tab.~\ref{tab:system} are due to the effect of varying $i$ within its allowed range.  

In shocked wind models, the properties of the stellar winds are relevant ingredients because the location of the intra-binary shock is determined by the dynamical pressure balance between the stellar and pulsar winds. For Be stars, the stellar wind is commonly described as being composed of a fast low-density polar wind and a slow dense equatorial disk wind~\citep{Waters-1988}. The disk properties of \bestar were studied by~\cite{Moritani-2015, Zamanov-2016} through optical spectroscopy. It was found that the disk size, estimated as the radius of the H$\alpha$ emitting region, is $0.85-1.4$~AU. The average disk radius of $1.12$~AU is indicated by dotted lines in Fig.~\ref{fig:system:orbits}, assuming that the disk and the orbit of the compact object are co-planar.

For the sets of observations reported here, both orbital solutions imply that the distance between the Be star and the pulsar is substantially larger than the disk size. Therefore, the influence of the disk on the shock formation in the colliding wind scenario is expected to be relevant only if the shock is located relatively close to the companion star. We will show, by the results of the model fitting, that the pulsar spin-down luminosity \edot required for the latter condition is not supported by our observations. Therefore, we will neglect the interaction with the disk in our model and only the polar-wind parameters will be defined.

The velocity profile of the polar wind is approximately described by $\vw(r) = v_0 + (v_{\infty}-v_0)(1-\rbe/r)$~\citep{Waters-1988}, where typically $v_0\sim20$~km/s and $v_{\infty}\sim1000-2000$~km/s. Because $r\gg \rbe$ in the context of our study, we use the approximation $\vw = v_{\infty}$, assuming $\vw = 1500$~km/s. The mass loss rate \mdot of the wind in Be stars is only poorly constrained and it is usually assumed to be in the range $10^{-9}-10^{-8}$~\msunyr based on~\cite{Snow-1981,Waters-1988}. The impact of the \mdot uncertainties on the results of the model fitting is substantial since the relative location of the intra-binary shock is strongly affected by this parameter. Thus, we will adopt $\mdot=10^{-8.5}$~\msunyr as a reference value and the range $10^{-9}-10^{-8}$~\msunyr will be used to provide an estimation of its uncertainties. 

\section{Description of the Model}
\label{sec:model}

Our model is based on the assumption that the compact object is a pulsar of mass $\mpsr=1.4$~\msun. The collision between the pulsar and the stellar wind creates a termination shock which is assumed to accelerate cold electron pairs from the pulsar wind to high energies. Hence, non-thermal radiation emitted by the accelerated electrons located around the apex of the shock is believed to produce the observed X-ray and gamma-ray photons through synchrotron and inverse Compton scattering (ICS), respectively. The pulsar spin-down luminosity, \edot, is a central parameter of the model, since it drives the pulsar wind pressure as well as the $B$-field strength. It is important to note that the following fitting approach is only meant to describe the observations presented in this paper with a minimum number of assumptions. The application of the same ideas to observations taken at other points in the orbit would require to take into account further processes, in particular, the effect of the circumstellar disk on the shock formation.

\subsection{Shock Formation and the Pulsar Wind Magnetization}
\label{sec:model:shock}

By imposing hydrodynamic balance between the pulsar and the stellar wind, the distance of the shock apex to the pulsar is given by~\citep{Harding-1990, Tavani-1997, Ball-2000}
\begin{equation}
    \rsh = \frac{\sqrt{\eta}}{1+\sqrt{\eta}} D,
    \label{eq:rsh}
\end{equation}
where
\begin{equation}
    \eta = \frac{\edot}{\dot{M}v_{\mathrm{w}}c},
    \label{eq:eta}
\end{equation}
and $D$ is the distance between both objects. $\dot{M}$ and $v_{\mathrm{w}}$ are the mass loss rate and the wind velocity assumed for the polar component of the stellar wind. The circumstellar disk component of the wind is neglected here because the distance between the objects is much larger than the size of the disk as estimated by~\cite{Moritani-2015, Zamanov-2016}. The influence of the disk, if present, would  move the shock position closer to the pulsar, changing the properties of the non-thermal emission. The same effect would be observed by assuming different values of $\dot{M}$, while still neglecting the disk component of the wind. Thus, in this model, the effect of disk interactions on the emission features are estimated by changing $\dot{M}$ within its uncertainties.

The particle acceleration, and consequently, non-thermal emission, is assumed to occur in a relatively small region around the shock apex, and therefore, the shock morphology will be neglected. This assumption is supported by the results of hydrodynamical simulations of the gamma-ray binary LS 5039~\citep{Dubus-2015}.

The relativistic electron pairs from the pulsar wind are accelerated at the termination shock, forming the high-energy electron population, upstream to the shock that is responsible for the non-thermal radiation. Following~\cite{Kennel-1984a, Kennel-1984b}, the $B$-field upstream of the shock is given by
\begin{equation}
    B = \sqrt{ \frac{\edot\sigma}{\rsh c (1+\sigma)} \left( 1 + \frac{1}{u^2}\right)},
    \label{eq:b}
\end{equation}
where $u$ is the radial four-velocity of the wind downstream from the shock, given by
\begin{equation}
\begin{split}
    u^2 = & \frac{8\sigma^2+10\sigma+1}{16(\sigma+1)}\\
    & + \frac{\left[64\sigma^2(\sigma+1)^2+20\sigma(\sigma+1)+1\right]^{1/2}}{16(\sigma+1)}.
    \label{eq:u}
\end{split}
\end{equation}
The magnetization of the wind $\sigma$ is commonly assumed to depend on the distance from the pulsar, as $\sigma \propto \rsh^{-\alpha}$ with $\alpha\sim 0.5-2.0$~\citep{Kong-2012, Takata-2017}.

\subsection{Energy Spectrum of the High-Energy Electron Population}
\label{sec:model:spec}

It is commonly assumed that the unshocked electron pairs from the pulsar wind are accelerated to a power-law energy distribution in the termination shock. After these electrons are injected into the downstream post-shock flow, radiative energy losses may change the energy spectrum, creating features that depart from the original power-law shape. The energy spectrum can be described by a broken power-law with an exponential cutoff. While the break is the result of the transition from being dominated by ICS, in the Klein-Nishima regime, to being dominated by synchrotron losses~\citep{Moderski-2005}, the cutoff is the result of the maximum acceleration energy provided by the confinement power of the $B$-field. Moreover, adiabatic energy losses may also play a relevant role in shaping the electron spectrum~\citep{Zabalza-2011}.        

The electron energy range relevant to describe our observations extends from the electrons responsible for producing the lowest energy X-ray photons (${\sim}3$ keV) through synchrotron emission, up to the ones responsible for producing the highest-energy gamma-ray photons (${\sim}3$ TeV) through ICS. At the upper bound, because the ICS occurs in the Klein-Nishima regime, the contribution from electrons with energy higher than the gamma-ray photons' is negligible, which implies that the maximum relevant electron energy is $E_{e,\text{max}}\lesssim 5$ TeV. At the lower bound, the energy of the electrons that contribute to the production of X-ray photons of energy \esyn is given by $E_e = \sqrt{\esyn m_e^3/eB}$. Since typical $B$ values found in gamma-ray binaries are in the range of $0.1-5$ G~\citep{Dubus-2013}, the minimum relevant electron energy is $E_{e, \text{min}} \approx 0.1$ TeV.

The maximum acceleration energy is estimated by balancing the acceleration and the energy-loss time scales, where the latter is dominated by the synchrotron losses at the highest energies~\citep{Khangulyan-2008}.  For typical parameters found in gamma-ray binaries, we determined that $E_{e, \text{max}} > 10$ TeV~\citep{Zabalza-2011, Dubus-2013}. 
The break energy, on the other hand, is expected to be ${\sim}0.1-0.5$ TeV for typical conditions during our observations of \src, which translates into a break in the X-ray synchrotron spectrum. Since the SED derived from our X-ray observations does not show any indication of deviation from a single power law, we assume here that the electron energy break appears at $E_{e} < 0.1$ TeV. Therefore, the electron spectrum will be assumed to follow a single power law of the form $\text{d}N_e/\text{d}E_e = \norm \left(E_e/1\;\text{TeV} \right)^{\slope}$ in the energy range $0.1-5$ TeV. The adiabatic energy losses do not affect this assumption because it is only expected to change the normalization and the slope of the spectrum, without creating any features. Thus, our model relies only on minimal assumptions about the electron spectrum, which is only possible due to the large overlap between the energies of the electrons responsible for producing hard X-rays through synchrotron emission and TeV gamma rays through ICS emission.        

\subsection{Radiative Processes}
\label{sec:model:rad}

The population of high-energy electrons accelerated at the termination shock is assumed to radiate through synchrotron and ICS, producing the X-ray and gamma-ray photons observed at Earth, respectively. The $B$-field responsible for the synchrotron emission is calculated following Eq.~\ref{eq:b}. The seed photon field for the ICS is composed of the thermal photons radiated by the companion star, with a density given by
\begin{equation}
    U = \frac{\sigma_{\text{SB}}\tbe^4 \rbe^2 }{c (D-\rsh)^2},
    \label{eq:density}    
\end{equation}
where $\sigma_{\text{SB}}$ is the Stefan-Boltzmann constant. We also account for the anisotropic nature of the ICS by calculating the photon scattering angle (\thetaic) from the geometry given by the orbital solutions.

Because of the relatively dense field provided by the stellar photons, the impact of pair-production absorption of gamma-rays~\citep{Gould-1967} becomes relevant. For each evaluation of the model, we calculate the optical depth ($\tau_{\gamma\gamma}$) as a function of the gamma-ray photon energy, accounting for the corresponding position of the termination shock and geometry of the system (see e.g.~\cite{Dubus-2006, Sierpowska-Bartosik-2008, Sushch-2017} for the details of the $\tau_{\gamma\gamma}$ calculation). The gamma-ray spectrum expected to be observed is then attenuated by a factor of $e^{-\tau_{\gamma\gamma}}$. The impact of pair-production absorption is the strongest at the low end of the observed gamma-ray spectrum (${\sim}0.2$ TeV), and it is larger for the~\cite{Casares-2012} orbital solution because of the longer path of the gamma-ray photons within the stellar photon field.  

\subsection{Model Fitting}
\label{sec:model:fit}

\begin{figure*}
\plottwo{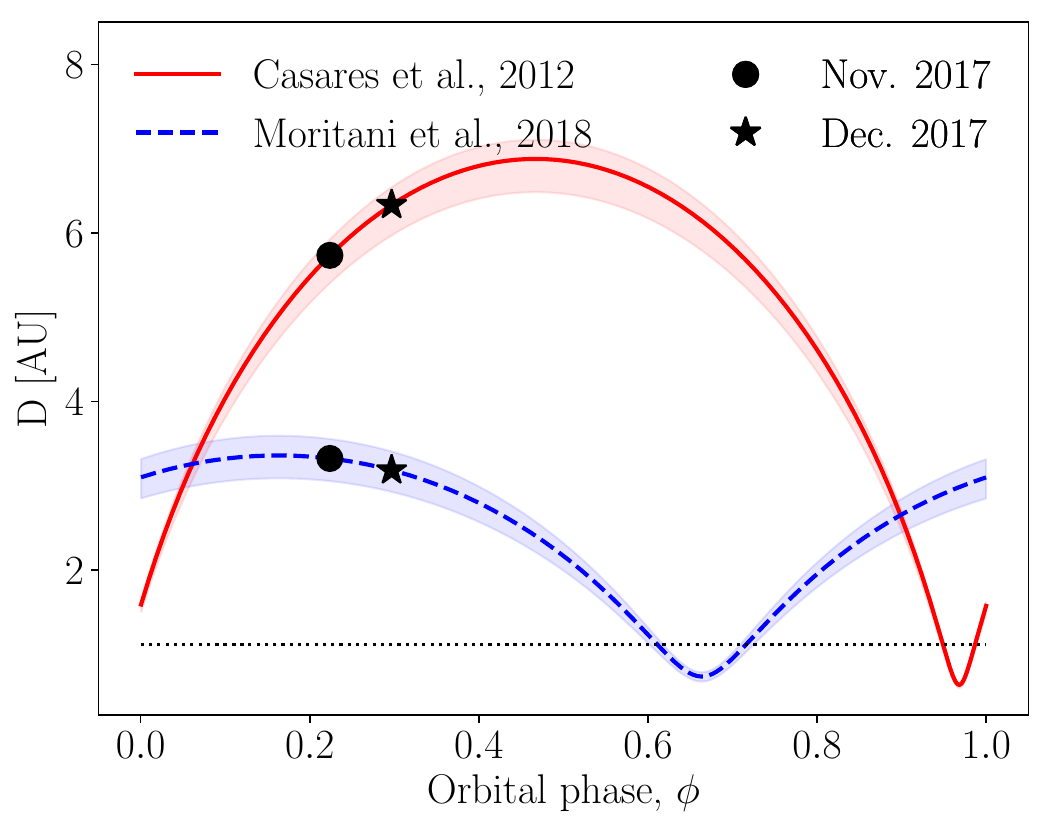}{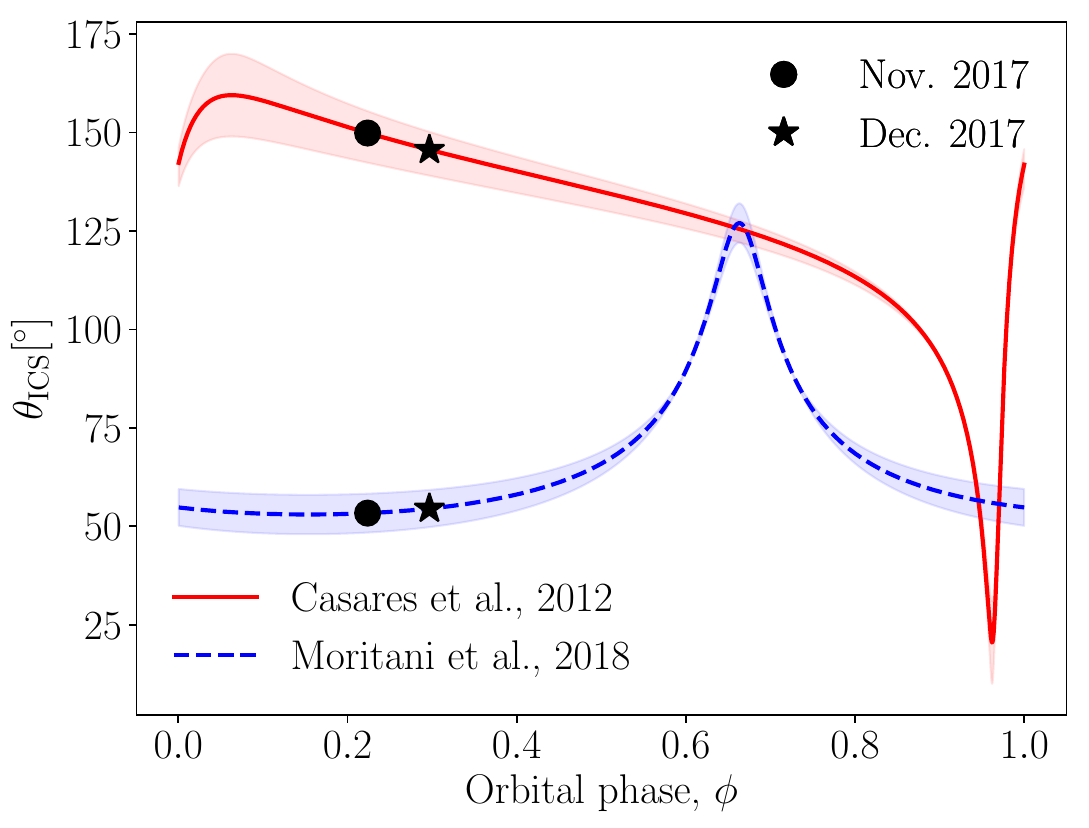}
\caption{\label{fig:system:distance}
Distance between the Be star and the pulsar (left) and ICS angle (right) derived from the orbital solutions by~\cite{Casares-2012} and~\cite{Moritani-2018}. The black markers indicate our two sets of observations, the dashed black line in the left panel indicates the estimated size of the cimcumstellar disk, and the colored bands indicate the uncertainties due to the system's inclination (see text for details).}
\end{figure*}

The model described above was fitted to the SED derived from both \nustar\ and VERITAS observations. In what follows, the two observation sets will be labelled by the lower indices $0$ and $1$, for the November and December 2017 observations, respectively. From the orbital solutions described in Sec.~\ref{sec:system}, we calculated the distance between the objects ($D$) and the ICS angle (\thetaic), which are shown in Fig~\ref{fig:system:distance} as a function of the system's phase. Assumptions are made regarding the stellar wind properties \mdot and \vw. The impact of the uncertainties on \mdot will be estimated by repeating the procedure with different values. 

The remaining parameters of the model are the pulsar spin-down luminosity (\edot), the pulsar wind magnetization at termination shock for both periods ($\sigma_0$ and $\sigma_1$), and the parameters of the electron spectrum for both periods (\normi{0}, \slopei{0}, \normi{1} and \slopei{1}). The dependence of $\sigma$ on \rsh is assumed to constrain $\sigma_0$ and $\sigma_1$ and reduce them to a single parameter. Here, we maintained $\sigma_0$ as a free parameter of the model and we calculated $\sigma_1=\sigma_0 \left(\frac{\rshi{0}}{\rshi{1}}\right)^\alpha$, with $\alpha=1$. Since the difference between \rshi{0} and \rshi{1} is relatively small for our observations, the impact of the choice of $\alpha=1$ is minimal.  

Regarding the electron spectrum, only \normi{0} and \normi{1} were treated as free parameters, while \slopei{0} and \slopei{1} were set to the values derived from the single power-law fit of the X-ray spectrum (see Sec.~\ref{sec:xray}). The values are $\slopei{0} = 1.77 \times 2 - 1 = 2.54$ and $\slopei{1} = 1.56\times 2 -1 = 2.12$. In the end, the model contains four free parameters: \edot, $\sigma_0$, \normi{0} and \normi{1}.

The model evaluation starts by inserting \edot into Eqs.~\ref{eq:rsh} and~\ref{eq:eta} to calculate \rshi{0} and \rshi{1} (using the assumptions of \mdot and \vw and $D_0$/$D_1$ given by the orbital solutions). With \rsh, the photon density $U$ of the ICS seed field and the optical depth $\tau_{\gamma\gamma}$ are calculated. In the following, the values of the $B$-field at the shock, $B_0$ and $B_1$, are calculated by inserting the $\sigma$ and \rsh values in Eq.~\ref{eq:b}. Finally, with the electron spectrum determined by \normi{0} and \normi{1}, the synchrotron and ICS spectra are calculated, using \thetaic from the orbital solutions for the ICS emission. 

The synchrotron and ICS emission were computed by following~\cite{Blumenthal-1970} using the Naima package~\citep{Zabalza-2015}. The SED fit is performed by means of a $\chi^2$ method. While \edot and $\sigma_0$ are scanned over a pre-defined grid, \normi{0} and \normi{1} are fitted by minimizing $\chi^2$ using Minuit framework~\citep{James-1975}.  

\section{Results}
\label{sec:results}

\begin{figure*}
\plottwo{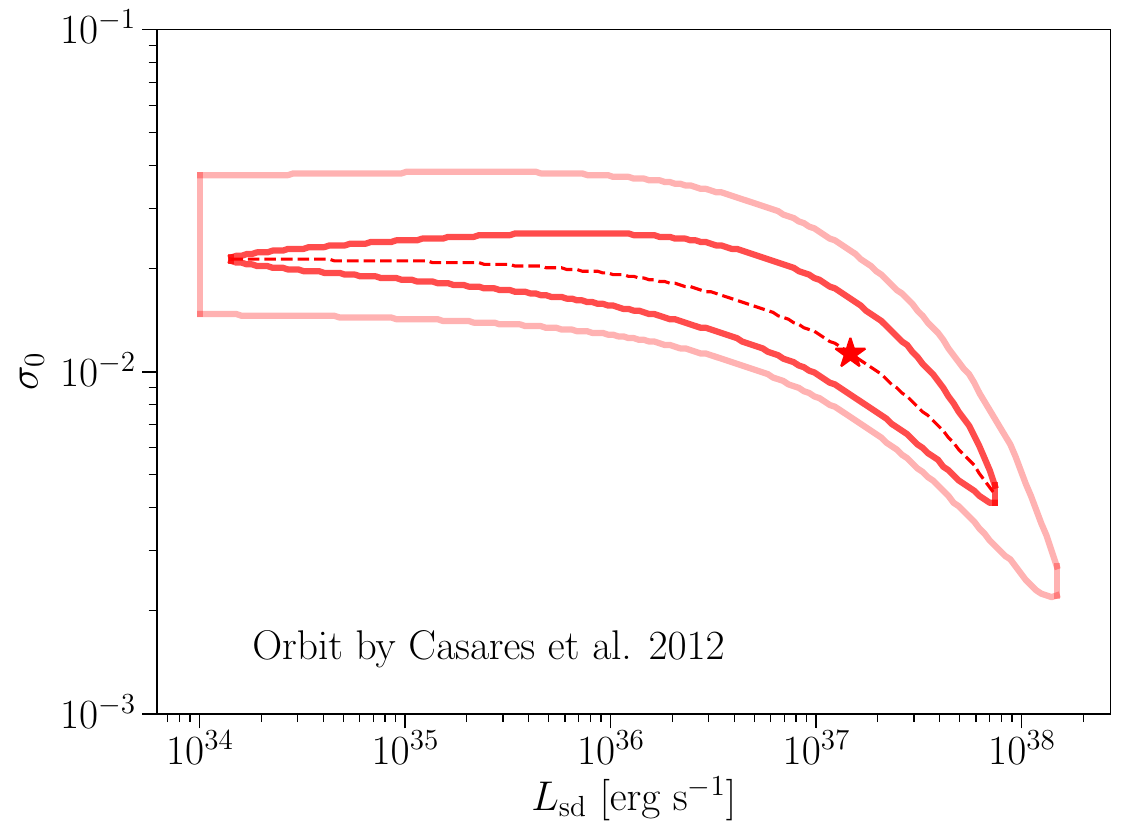}{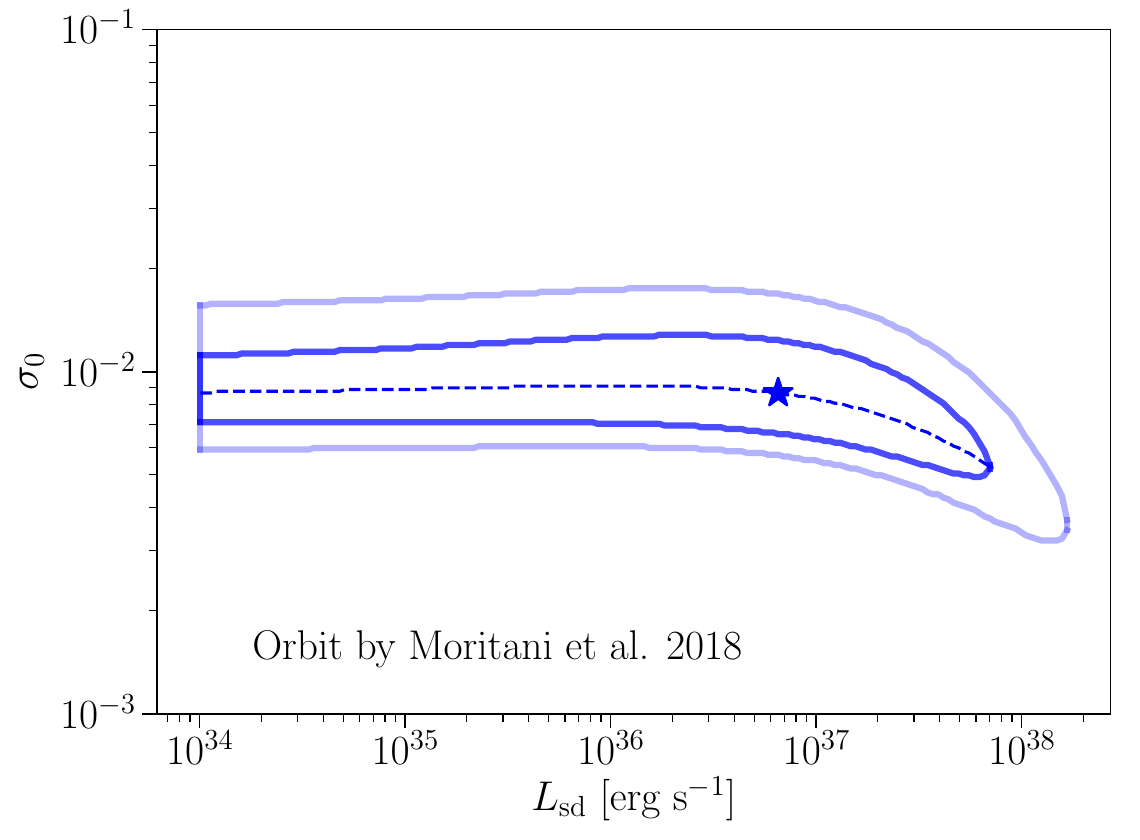}
\caption{\label{fig:results:solutions}
Results of the model fitting in the $\edot-\sigma_0$ plane for the sets of orbital parameters from~\cite{Casares-2012} (left) and~\cite{Moritani-2018} (right). The best solution is indicated by a star while the 1 and $2\sigma$ regions are indicated by the darker and lighter continuous lines, respectively. The dashed lines show the values of $\sigma_0$ that minimize the $\chi^2$ for each value of \edot.}
\end{figure*}

\begin{figure*}
\plotone{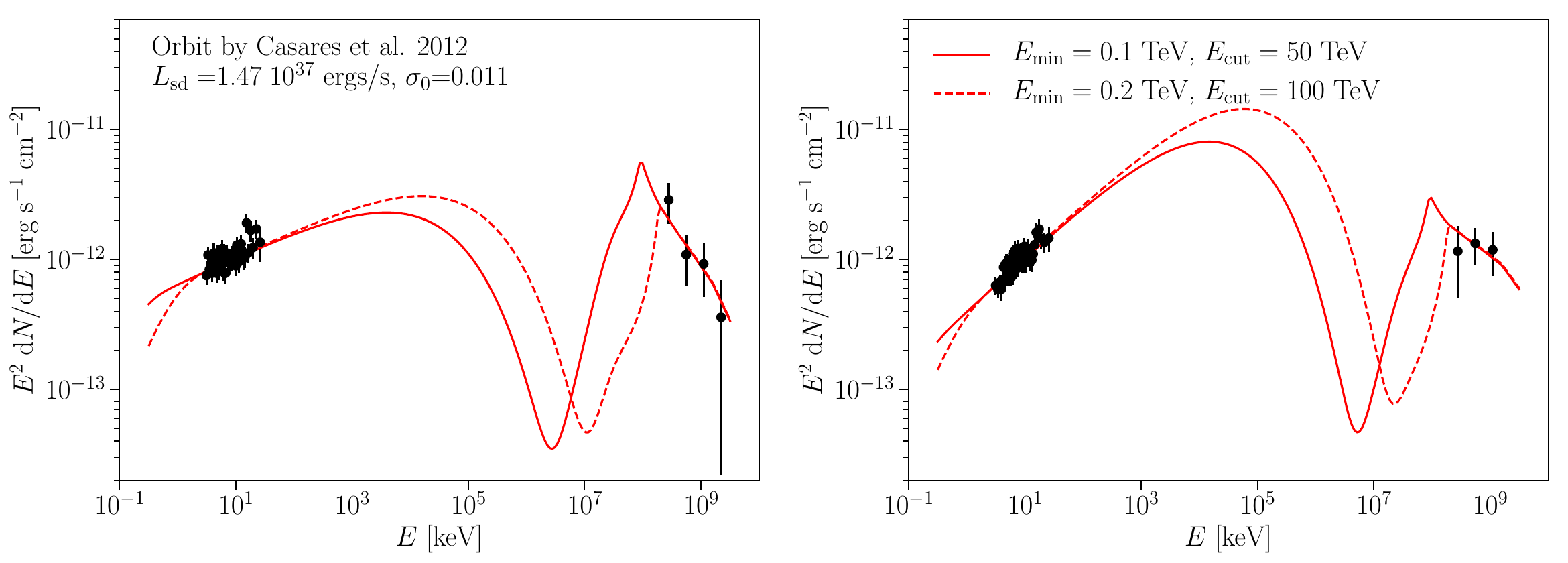}

\plotone{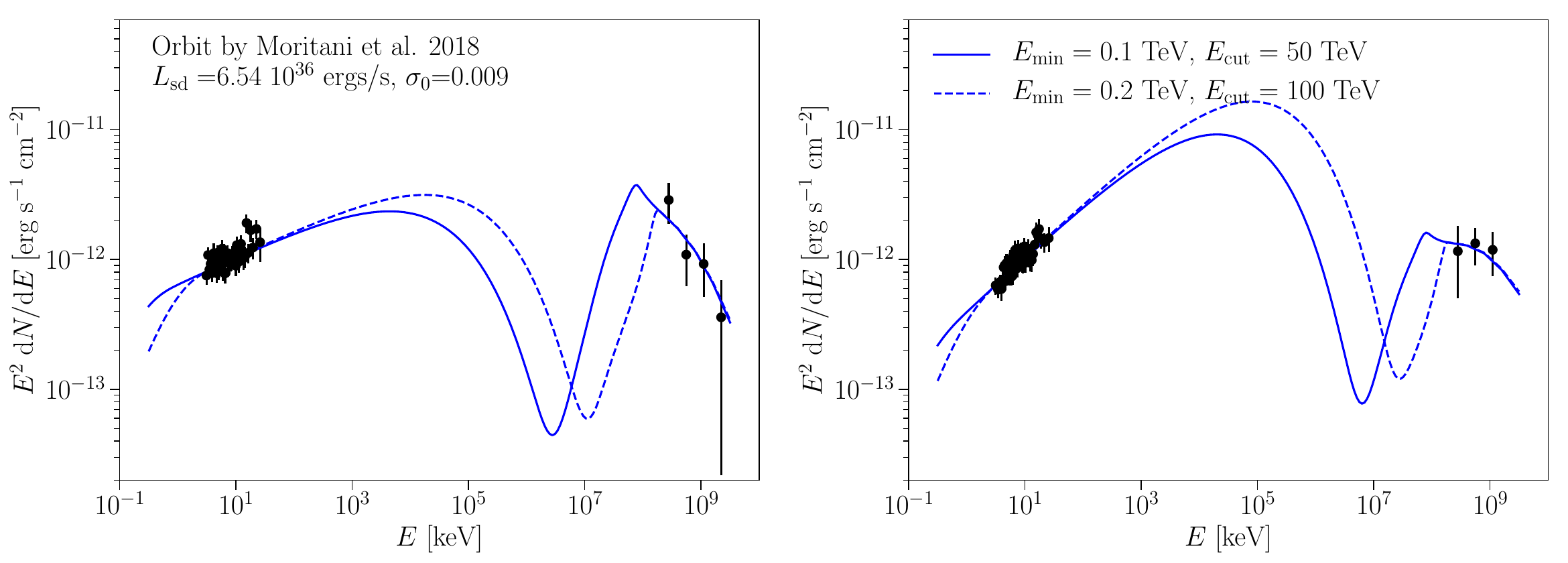}
\caption{\label{fig:results:sed}
SED data-model comparison assuming the best solution of the model fitting for the~\cite{Casares-2012} (upper) and~\cite{Moritani-2018} (lower) orbital solutions. The November and December 2017 observations are shown in the left and right panels, respectively. The different line styles indicate different assumptions of $E_{\text{min}}$ and $E_{\text{cut}}$.}
\end{figure*}

We show in Fig.~\ref{fig:results:solutions} the results of the fit for the \edot-$\sigma_0$ plane, indicating the 1 and 2$\sigma$ regions for both orbital solutions. The best solutions are indicated by stars and the corresponding $\chi^2/\text{dof}$ is $0.786$ for both sets of orbital parameters. The dashed lines indicate the $\sigma_0$ that minimizes the $\chi^2$ for each value of \edot. These curves will be labeled as \textit{best curves} in what follows. The SED model-data comparisons are shown in Fig.~\ref{fig:results:sed}, in which the best solution of the fit is used to evaluate the model. To illustrate the SED throughout all the energy bands, the electron spectrum was assumed to be a power-law, starting at $E_{\text{min}}$, with an exponential cutoff characterized by $E_{\text{cut}}$. Although different values of $E_{\text{min}}$ and $E_{\text{cut}}$ are illustrated in Fig.~\ref{fig:results:sed}, the model fitting does not depend on these parameters, as explained in Sec.~\ref{sec:model:spec}.



\begin{figure}
\plotone{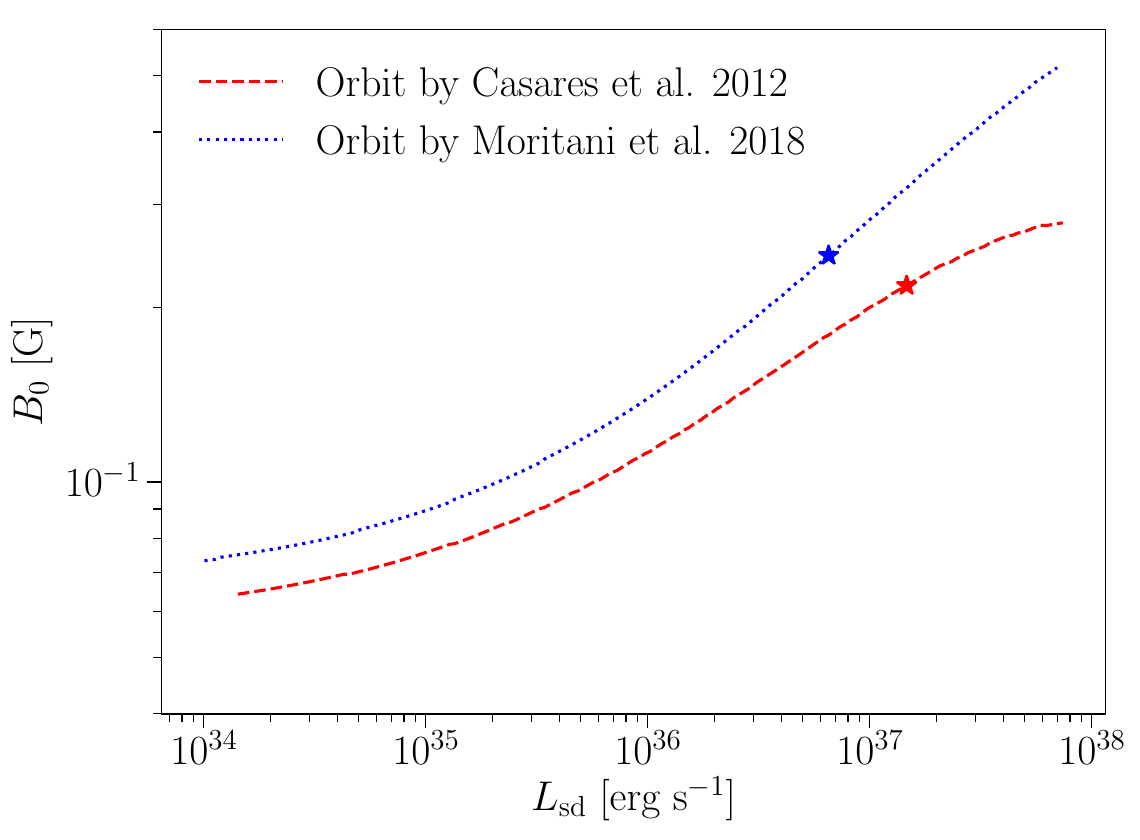}

\plotone{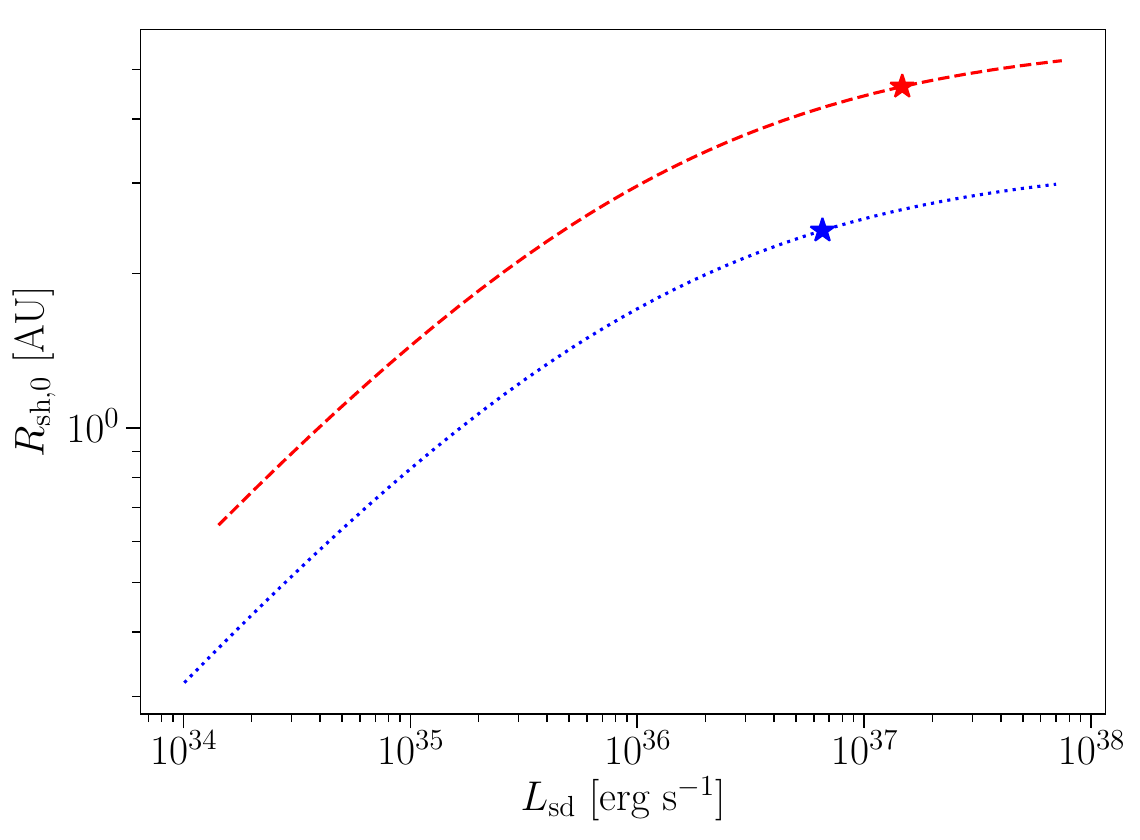}

\plotone{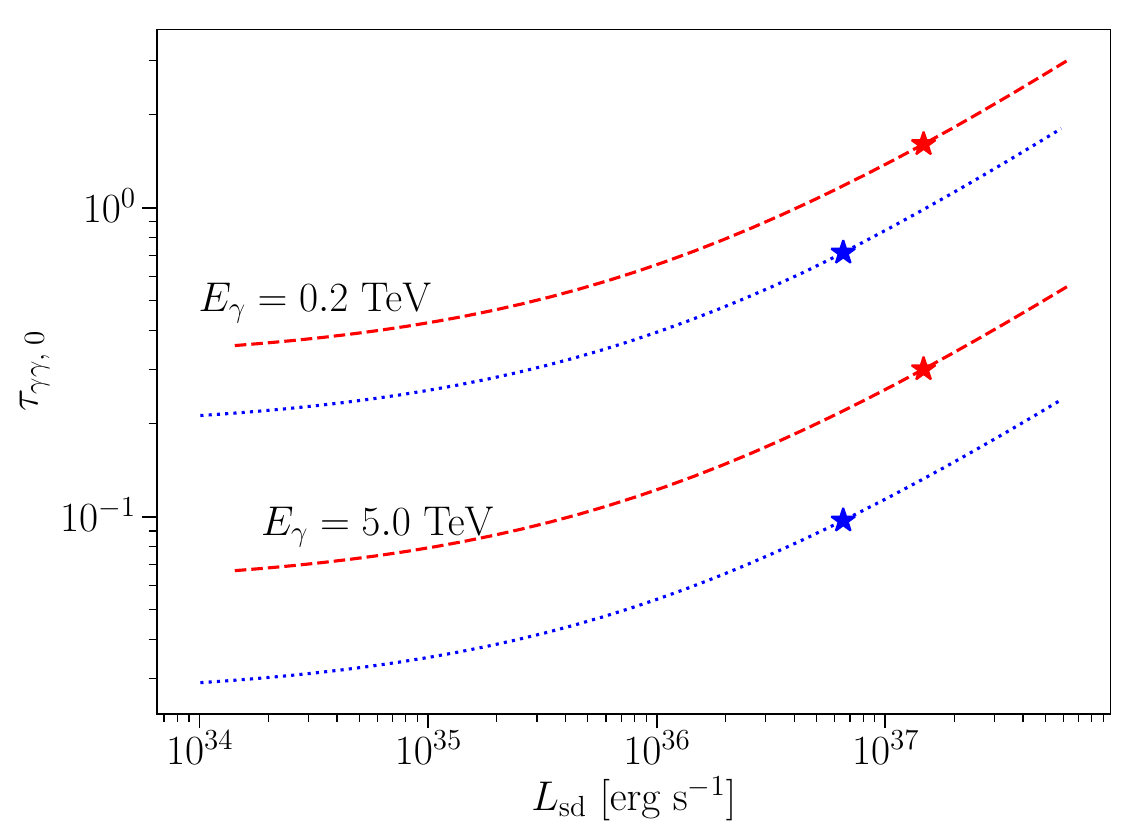}
\caption{\label{fig:results:bfield}
$B$-field (upper), shock distance to the pulsar (middle) and pair production optical depth (lower) for the earlier set of observations as a function of \edot. The lines represent the best curves and the stars show the best solutions. The optical depth $\tau_{\gamma\gamma,0}$ is shown for gamma-ray photon energies of $0.2$ (upper lines) and $5.0$ TeV (bottom lines).}
\end{figure}

The degeneracy observed between \edot and $\sigma_0$ implies that neither of these parameters can be individually constrained with our approach. Nevertheless, our results demonstrate that the observations can be consistently described by a pulsar wind model and indicate the region of the $\edot$-$\sigma_0$ space that allows for it. To further characterize the allowed solutions, we show in Fig.~\ref{fig:results:bfield} the $B$-field, \rsh and $\tau_{\gamma\gamma}$ for the November observation ($B_0$, \rshi{0} and $\tau_{\gamma\gamma, 0}$) as a function of \edot. The lines in Fig.~\ref{fig:results:bfield} correspond to the best curves shown in  Fig.~\ref{fig:results:solutions}. 

The estimated total luminosity of the full SED shown in Fig.~\ref{fig:results:sed} is in the range $10^{34}-10^{35}$ \lum. Estimations of the efficiency of non-thermal radiation in gamma-ray binaries are highly uncertain. For instance, \fermi observations show that nearly all the \edot is converted into non-thermal radiation in PSR~B1259-63 around the periastron passage~\citep{Abdo-2011}. \cite{Sierpowska-Bartosik-2008}, on the other hand, assume an efficiency of 0.01 for LS~5039. Therefore, the truncation of our results at $\edot=10^{34}$~\lum represents a conservative lower limit. 

The impact of the different system geometries provided by the two orbital solutions is minimal. The observations are equally well described by both sets of orbital parameters, with no indication of one of them being favored. While $\sigma_0$ for $\edot < 10^{37}$ \lum is $2-3$ times larger for the~\cite{Casares-2012} solution, the difference in $B_0$ amounts to only a factor of $0.6$ at the highest \edot regime.

At the high \edot regime of the solutions, the upper limit within $1\sigma$ is $\edot<\sn{7}{37}$ \lum for both sets of orbital parameters, which is consistent with expectations for very young pulsars. In this regime, the relatively strong pulsar wind moves the termination shock closer to the companion star (see Fig.~\ref{fig:results:bfield}, middle). For both orbital solutions, the termination shock is approximately halfway between the stars for the highest \edot values, although the absolute distances are different. The proximity of the shocked region to the Be star implies a larger density for the ICS seed-photon field, and consequently, enhanced ICS emission, but also stronger pair-production absorption. Since the ratio between ICS and synchrotron emission is constrained by the data, the $B$-field has to change accordingly to compensate for the ICS behavior, which is done by changing $\sigma_0$. In this regime, $B_0\approx 0.25$ and $\approx 0.5$ G for the \cite{Casares-2012} and \cite{Moritani-2018} solutions, respectively.

As the \edot of the solutions decreases, the termination shock moves closer to the pulsar, within a small fraction of the distance between the objects. At this regime, the $B$-field also decreases and reaches $B_0\approx0.06$ and $0.07$ G for the \cite{Casares-2012} and \cite{Moritani-2018} solutions, respectively. 

\begin{figure*}
\plottwo{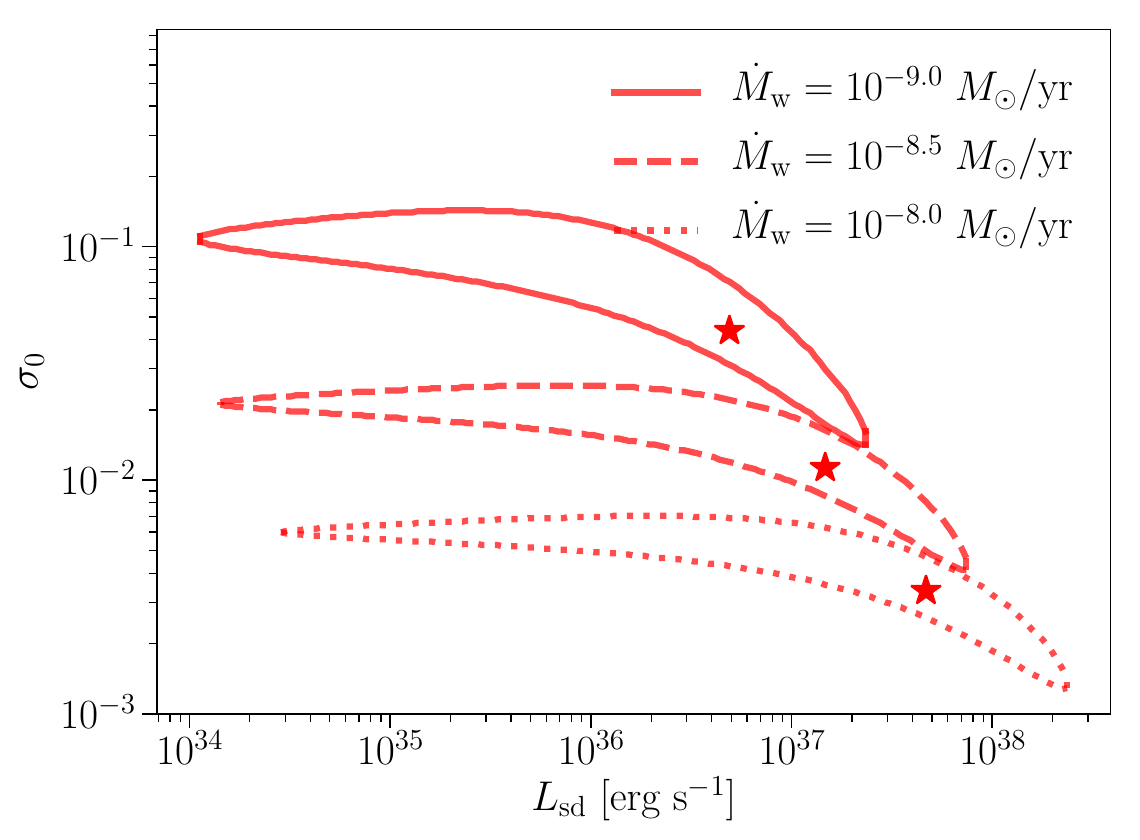}{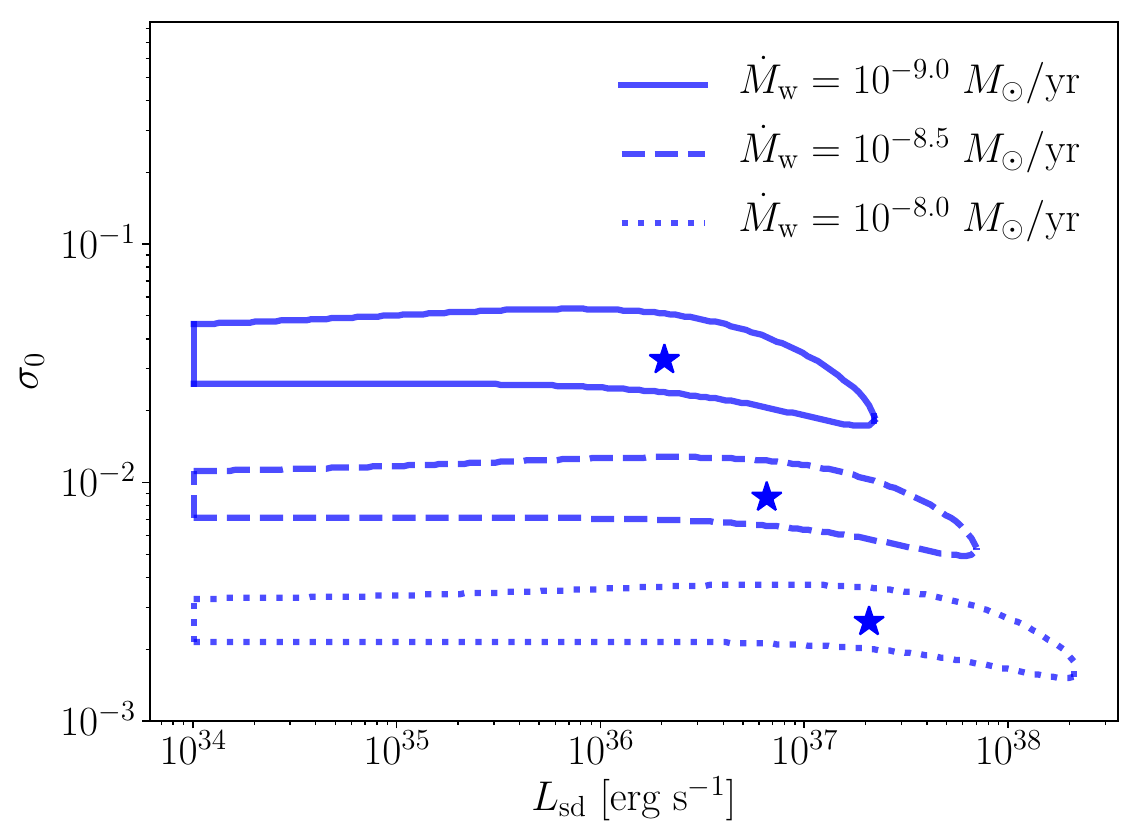}
\caption{\label{fig:results:mdot}
Results of the model fitting in the $\edot-\sigma_0$ plane for different values of \mdot within the range $10^{-9}-10^{-8}$~\msunyr.  The orbital solutions by~\cite{Casares-2012} and~\cite{Moritani-2018} are shown in the left and right panel, respectively. The lines indicate the $1\sigma$ region of the parameters.}
\end{figure*}

\begin{figure*}
\plottwo{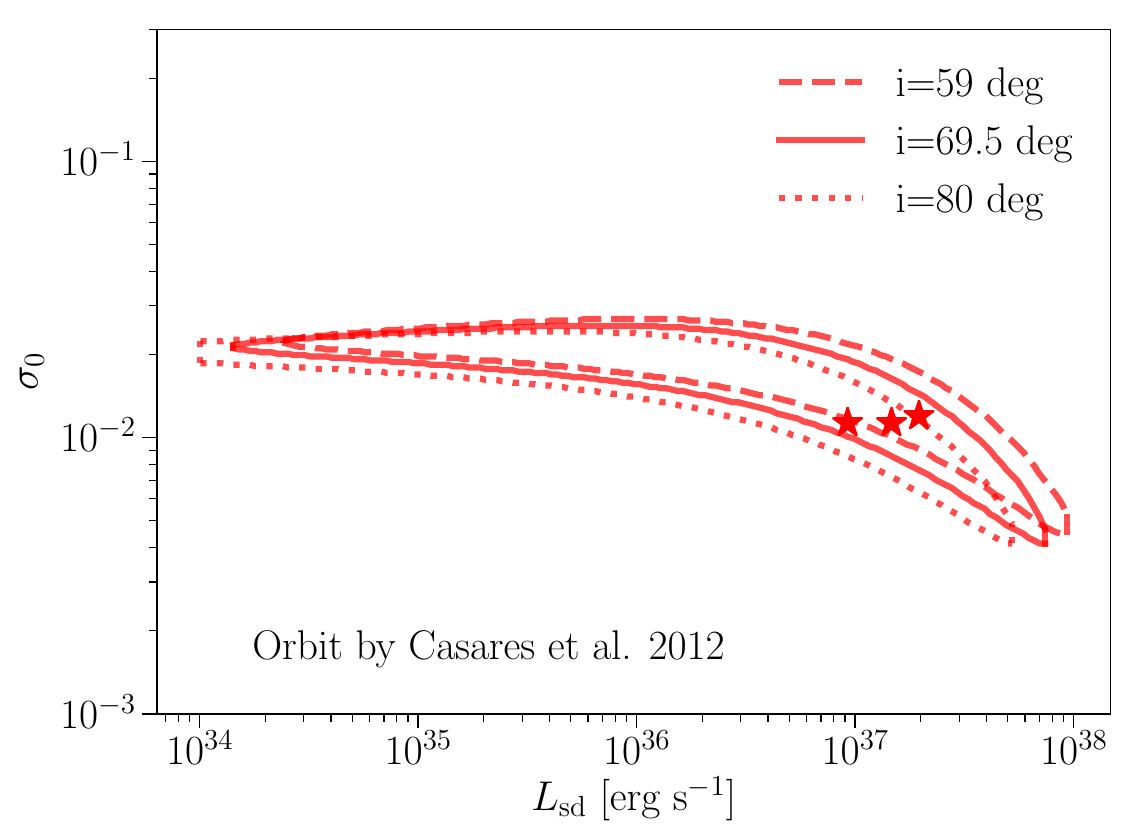}{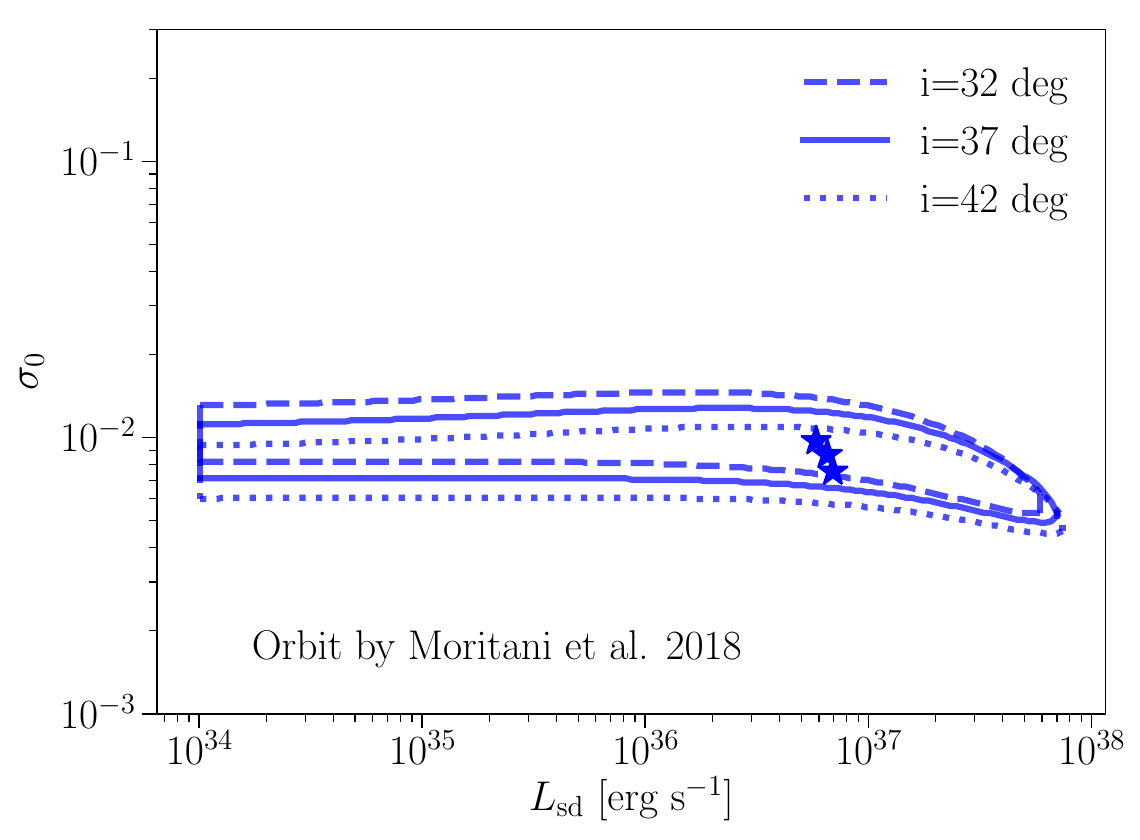}
\caption{\label{fig:results:inc}
Results of the model fitting in the $\edot-\sigma_0$ plane for different values of inclination as described in Sec.~\ref{sec:system}. The orbital solutions by~\cite{Casares-2012} and~\cite{Moritani-2018} are shown in the left and right panel, respectively. The lines indicate the $1\sigma$ region of the parameters.}
\end{figure*}

The impact of the uncertainties on the system's parameters was evaluated by varying the assumed values and repeating the fitting procedure. We found that the most relevant uncertainty is that of the mass loss rate of the stellar wind, \mdot. The results of the fit for $\mdot = 10^{-9.0}, 10^{-8.5}$ and $10^{-8.0}$~\msunyr are shown in Fig.~\ref{fig:results:mdot}. We also show the results in Fig.~\ref{fig:results:inc} by accounting for the uncertainties in the system's inclination. Note that as described in Sec.~\ref{sec:system}, the variations of the inclination also result in variations in the semi-major axis, as well as in the mass of the companion star. Apart from the \mdot and inclination, we also tested the robustness of the results considering the uncertainties in the eccentricity, $\omega$, \porb, and $\alpha$. In all cases the effect is similar to or slightly larger than that observed for the inclination. The uncertainty range considered for each of these parameters is indicated in Tab.~\ref{tab:system}.

The impact of the poorly known properties of the stellar wind, represented by the mass loss rate \mdot, is the most relevant source of uncertainty on the constraining power of the pulsar-wind magnetization from our approach. The large range of \mdot assumed here is sufficient to incorporate any effect due to the presence of the circumstellar disk, which has been neglected by the model assumptions.

\section{Summary and Discussions}
\label{sec:conclusions}

\begin{figure*}
\plotone{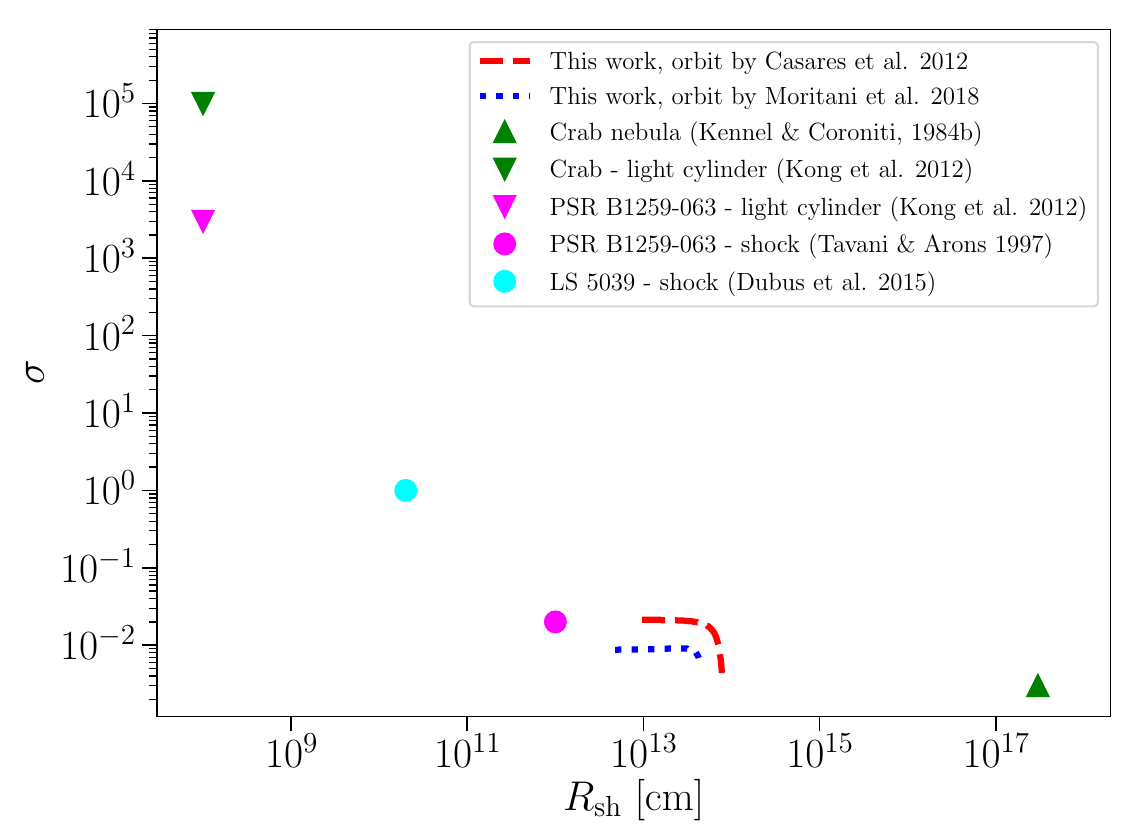}
\caption{\label{fig:results:magnetization}
Pulsar-wind magnetization \textit{vs} \rsh, obtained from the results of our model fitting, compared to a selection of theoretical and observational constraints (see text for details).}
\end{figure*}

We presented the results of two sets of combined observations of the gamma-ray binary \src, by \nustar\, in the hard X-ray band, and by VERITAS, in the TeV gamma-ray band. These observations correspond to the rise of the first peak observed in the X-ray and TeV light curves, at $\phi\approx0.22$ and $0.30$. The data provided by both instruments allows us to derive the combined SED for these two periods. The spectral and timing analysis performed on the \nustar\ observations show that: a) the spectra are well described by a single power law model with a spectral hardening observed between the two observations ($\Gamma$ going from $1.77\pm0.05$ to $1.56\pm0.05$), and b) no evidence of time variability, red noise, or pulsation has been found, consistent with observations of other gamma-ray binaries. 

The SED data is used to probe a model based on the pulsar wind scenario, in which the non-thermal emission is produced by high energy electrons accelerated at the termination shock formed by the interaction between pulsar and stellar wind. The applied model relies on a minimum number of assumptions which are sufficient for the description of the observations presented in this paper. The description of further observations would require to expand the number of assumptions and the complexity of the model.

The model fitting benefited from two characteristics of our observations. Firstly, for both orbital solutions, the observations correspond to periods in which the distance between the Be star and the compact object is sufficiently large so that the influence of the circumstellar disk is negligible, reducing the number of assumptions required to describe the properties of the stellar wind. Secondly, in systems like gamma-ray binaries, there is a large overlap between the energy range of the electrons responsible for producing hard X-rays and very-high-energy gamma-rays (through synchrotron and ICS, respectively), which allows us to simplify the assumptions about the shape of the electron spectrum to a single power law description, with only two parameters. 

The results of the model fitting show the regions of the $\edot$-$\sigma$ plane that are allowed by our data. We find an upper limit of $\edot<\sn{7}{37}$ \lum within $1\sigma$, independent of the orbital solution, which is consistent with expectations of young pulsars. The $\sigma$ parameter is constrained to be $0.003-0.03$ at the location of the shock. Constraints on $\sigma$ are particularly important for understanding the physical process behind the transport of energy from the rotation powered pulsar to the surrounding medium, which is still a subject of intense discussions~\citep{Arons-2002, Kirk-2009}. While theoretical models predict that at the light cylinder the pulsar wind is dominated by Poynting energy ($\sigma_{L}\gg1$), observations of the Crab Nebula constrain $\sigma$ at much larger distances to be kinetic particle dominated ($\sigma_{N}\ll1$). The transition between these two regimes is not well described within the current theoretical framework, originating the so-called ``$\sigma$ problem''. In gamma-ray binaries, the pulsar wind termination is typically located at intermediate distances between the light cylinder and the termination shock in pulsar wind nebulae, which makes these systems interesting in this context.


In Fig.\ref{fig:results:magnetization} we compare the $\sigma$ {\it vs} \rsh of the best curves obtained from our fitting with the following selection of results: a) at the light cylinder ($\rsh\approx10^{8}$ cm), \cite{Kong-2012} estimates theoretically that $\sigma\approx\sn{1}{5}$ and $\sn{8}{3}$ for the Crab Pulsar and for PSR~J1259-63, respectively, b) \cite{Tavani-1997} assume $\sigma=0.02$ at the termination shock to describe observations of PSR~J1259-63 around the periastron, c) \cite{Dubus-2015} constrains $\sigma\approx1$ at the termination shock to describe observations of the system LS~5039 using a numerical hydrodynamical model and d) the classical result by \cite{Kennel-1984b} that constrains $\sigma\approx0.003$ at $\rsh\approx10^{17}$ cm for the Crab Nebula. The comparison indicates that all systems follow a similar trend of $\sigma$ decreasing with \rsh. The consistency observed between our results and other systems provides further support to the pulsar scenario of \src.

\section*{Acknowledgements}
This work used data from the \nustar\ mission, a project led by the California Institute of Technology, managed by the Jet Propulsion Laboratory, and funded by NASA. We made use of the \nustar\  Data Analysis Software (NuSTARDAS) jointly developed by the ASI Science Data Center (ASDC, Italy) and the California Institute of Technology (USA).

This research is supported by grants from the U.S. Department of Energy Office of Science, the U.S. National Science Foundation and the Smithsonian Institution, and by NSERC in Canada. We acknowledge the excellent work of the technical support staff at the Fred Lawrence Whipple Observatory and at the collaborating institutions in the construction and operation of the VERITAS instrument.

\facilities{NuSTAR, VERITAS}

\software{HEAsoft (v6.22), \texttt{Stingray} \citep{Stingray-2016}, NuSTARDAS (v1.8.0), XSPEC (v12.9.1; \cite{Arnaud-1996}), Naima \citep{Zabalza-2015}}

\end{document}